\documentclass{article}
\usepackage{ijcai26}

\usepackage{times}
\usepackage{soul}
\usepackage{url}
\usepackage[hidelinks]{hyperref}
\usepackage[utf8]{inputenc}
\usepackage[small]{caption}
\usepackage{graphicx}
\usepackage{amsmath}
\usepackage{amsthm}
\usepackage{booktabs}
\usepackage{algorithm}
\usepackage{algorithmic}
\usepackage[switch]{lineno}
\usepackage{natbib}
\usepackage{amssymb} 
\usepackage{graphicx}
\usepackage{subcaption}
\usepackage{wrapfig}
\usepackage{enumitem}

\title{DiffLOB: Diffusion Models for Counterfactual Generation in Limit Order Books}

\author{
Zhuohan Wang$^1$
\and
Carmine Ventre$^2$\and
\affiliations
$^1$King's College London
\emails
\{zhuohan.wang, carmine.ventre\}@kcl.ac.uk
}

\begin{document}

\maketitle

\begin{abstract}
    Modern generative models for limit order books (LOBs) can reproduce realistic market dynamics, but remain fundamentally passive: they either model what typically happens without accounting for hypothetical future market conditions, or they require interaction with another agent to explore alternative outcomes. This limits their usefulness for stress testing, scenario analysis, and decision-making. We propose \textbf{DiffLOB}, a regime-conditioned \textbf{Diff}usion model for controllable and counterfactual generation of \textbf{LOB} trajectories. DiffLOB explicitly conditions the generative process on future market regimes--including trend, volatility, liquidity, and order-flow imbalance, which enables the model to answer counterfactual queries of the form: ``If the future market regime were X instead of Y, how would the limit order book evolve?'' Our systematic evaluation framework for counterfactual LOB generation consists of three criteria: (1) \textit{Controllable Realism}, measuring how well generated trajectories can reproduce marginal distributions, temporal dependence structure and regime variables; (2) \textit{Counterfactual validity}, testing whether interventions on future regimes induce consistent changes in the generated LOB dynamics; (3) \textit{Counterfactual usefulness}, assessing whether synthetic counterfactual trajectories improve downstream prediction of future market regimes. 
\end{abstract}

\section{Introduction}


Modern financial markets are 
driven by automated trading systems operating on limit order books (LOBs) ~\citep{gould2013limit}. A LOB can be thought of as two priority queues collecting buy and sell orders of market participants. Each order is comprised of bid/ask price (how much one is willing to pay/get for each stock) and volume (how many stocks one wants to trade). Small changes in trend, liquidity, volatility, or order flow balance of the LOB can lead to drastically different market outcomes. As a result, stress testing, strategy design, and market surveillance all require not only accurate models of how markets behave on average, but also the ability to simulate how the market would become under hypothetical future conditions, such as fast trend change, liquidity shocks, volatility spikes, or imbalanced order flow. 

However, existing approaches to LOB modeling exhibit fundamental limitations in answering counterfactual questions. Supervised LOB predictors typically focus on forecasting specific future variables, such as mid-price returns~\citep{briola2025deep, briola2025hlob}, but they do not generate full market trajectories and therefore cannot simulate how the order book would evolve over time under alternative future conditions. Recent generative models for LOBs are capable of producing highly realistic samples when conditioned on historical observations~\citep{nagy2023generative, coletta2021towards, limars}. However, to obtain counterfactual outcomes, these models usually rely on explicit interactions with trading agents. As a result, the  counterfactual trajectories generated are not guaranteed as a desired future state of the market.

To address this gap, we propose DiffLOB\footnote{Our code is available at \url{https://github.com/ZhuoHan1998/DiffLOB}.}, a diffusion model-based framework specifically designed for the counterfactual generation of LOB snapshots. The proposed framework offers three main contributions:

\begin{itemize}[nosep]
    \item[${(1)}$] We formulate counterfactual generation of LOBs as a conditional generative modeling problem and explicitly introduce future market regimes as control variables to enable controllable and counterfactual generation. We develop a novel diffusion-based architecture that conditions on past LOB trajectories, time-of-day information, and four future regime variables--trend, volatility, liquidity, and order-flow imbalance. The proposed model effectively captures key financial statistics and consistently outperforms strong baseline models across a range of evaluation metrics.
    
    \item[${(2)}$] We empirically demonstrate that interventions on future market regimes induce consistent and interpretable changes in the generated LOB dynamics. Under extreme and hypothetical regime conditions, DiffLOB produces trajectories whose statistical properties align with the imposed regimes and closely match real markets observed under comparable conditions.
    
    \item[${(3)}$] We show that synthetic counterfactual trajectories generated by DiffLOB provide meaningful additional information for downstream tasks. In particular, augmenting real data with counterfactual samples consistently improves the performance of models for future market regime prediction under extreme conditions, demonstrating the practical value of counterfactual LOB generation.
\end{itemize}

\section{Related Work}

\noindent \textbf{Generative Modelling in LOBs.} 
A growing body of work applies generative models to LOBs with the goal of reproducing realistic market microstructure dynamics. Autoregressive approaches have been used to model LOB dynamics at the event level.
\citet{hultin2023generative} decompose the joint distribution of LOB transitions into conditional components using recurrent neural networks,
while \citet{nagy2023generative} propose an end-to-end autoregressive model based on structured state-space models that tokenizes message streams.
GAN-based approaches, on the other hand, employ adversarial training to synthesize LOB data with high visual and statistical realism.  ~\citet{coletta2021towards} propose a Conditional GAN framework that reacts to current market states and allows agent interaction within a simulation, showing enhanced realism and responsiveness. Recent work has also explored diffusion-based approaches for limit order book simulation. TRADES~\citep{berti2025trades} develops a diffusion-based market simulator that generates order-level LOB dynamics and supports counterfactual analysis through interaction with trading agents. DiffVolume~\citep{wang2025diffvolume} apply diffusion models to generate high-dimensional LOB volume snapshots across multiple price levels without capturing full LOB trajectories. ~\citet{limars} create an order-level generative foundation model for downstream forecasting, risk detection and financial analysis. Despite their success in realism, existing generative LOB models primarily learn the observational distribution of market dynamics. They are not designed to explicitly intervene on future market regimes, and therefore cannot generate counterfactual and complete LOB trajectories corresponding to hypothetical future conditions.

\paragraph{Diffusion Models for Financial Time Series.} Diffusion models are a class of generative models inspired by thermodynamic diffusion processes, where data are progressively perturbed with Gaussian noise and generated by learning to reverse this process \citep{sohl2015deep, ho2020denoising}. 
This formulation is closely connected to score-based generative modeling, which learns the gradient of the data density via denoising score matching \citep{vincent2011connection, song2019generative}, and has been unified under a stochastic differential equation (SDE) framework \citep{song2020score}. 
Building on these advances, diffusion models have been increasingly applied to financial time series. \citet{koa2023diffusion} employ diffusion models for multi-step stock price prediction,
while \citet{wang2024financial} apply diffusion-based denoising techniques to financial time series.
More recently, \citet{tanaka2025cofindiff} propose a controllable diffusion framework for financial time series generation and introduce normalization-based techniques to enhance controllability \citep{hashimoto2025norm}.

\section{Methodology}

\subsection{Problem Formulation}

Let $x_t \in \mathbb{R}^{K \times C}$ denote the state of LOB at time $t$, where $K$ denotes the number of price levels (e.g., 10 ask price levels and 10 bid price levels) and $C$ represents the feature dimension (e.g., price and volume).
Given a historical LOB trajectory $x_{1:t}$, our goal is to generate a future trajectory $x_{t+1:t+\tau}$. We introduce future market regimes as controlling variables to enable controllable and counterfactual generation.
Specifically, we define a set of future regime variables
\begin{equation*}
\resizebox{0.91\hsize}{!}{$
\begin{split}
c_{t+1:t+\tau} &= \big(c^{\text{trend}}_{t+1:t+\tau}, c^{\text{vol}}_{t+1:t+\tau}, c^{\text{liq}}_{t+1:t+\tau}, c^{\text{imb}}_{t+1:t+\tau}\big), \\
\text{where} \ &
\{c^{\text{trend}}_{t+1:t+\tau}, c^{\text{vol}}_{t+1:t+\tau} \}\in \mathbb{R}, \ 
\{c^{\text{liq}}_{t+1:t+\tau}, \ c^{\text{imb}}_{t+1:t+\tau} \}\in \mathbb{R}^{\tau}
\label{eq:condition-formulation}
\end{split}
$}
\end{equation*}
corresponding to trend, volatility, liquidity, and order-flow imbalance, respectively.
These variables characterize the macroscopic state of the market over the future horizon and are treated as \emph{intervenable control variables} rather than observed labels. Our objective is to model the distribution
\begin{equation}
p\big(x_{t+1:t+\tau} \mid x_{1:t}, c_{t+1:t+\tau}\big),
\label{eq:problem-formulation}
\end{equation}
which enables counterfactual queries of the form:
\emph{``How would the limit order book evolve if the future market regime conditions were different?''}
By explicitly conditioning on future regimes, DiffLOB decouples market dynamics from specific agent behaviors and provides direct control over hypothetical future market conditions. 

We adopt a denoising diffusion probabilistic model (DDPM) ~\citep{ho2020denoising} to represent the conditional distribution of future LOB trajectories. In the forward process, Gaussian noise is progressively added to the target trajectory $x_{t+1:t+\tau}$ over a sequence of diffusion steps. A neural network is trained to reverse this process by predicting the injected noise at each step, thereby learning to generate samples from the data distribution. Conditions on historical trajectories and future market regimes are incorporated through the denoising network, allowing the diffusion process to generate future LOB trajectories consistent with specified regimes.
We follow standard score-based diffusion modelling and refer the reader to Appendix \ref{sec:theory} for full technical details.

\subsection{DiffLOB Architecture}
\label{subsec:difflob-architecture}

\begin{figure*}[t]  
    \centering  
    \includegraphics[width=\textwidth]{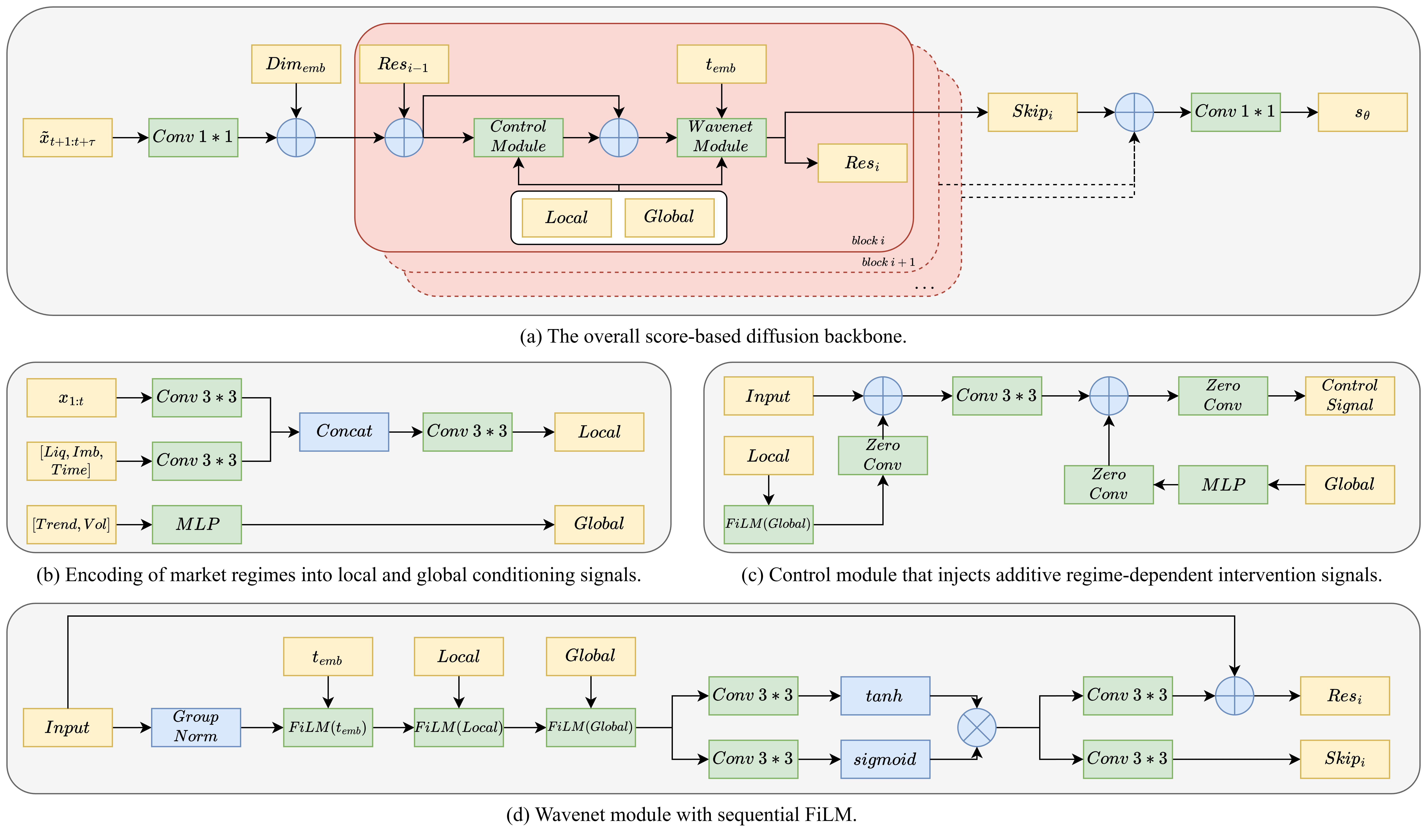}
    \caption{Illustration of DiffLOB Architecture.}
    \label{fig:architecture}  
\end{figure*}

An overview of DiffLOB architecture is shown in Figure~\ref{fig:architecture}.

\paragraph{Overall Backbone.}
As illustrated in Figure~\ref{fig:architecture}(a), DiffLOB adopts stacked Wavenet-style residual blocks ~\citep{van2016wavenet}. The backbone models the structural and temporal dynamics of LOB price and volume evolution, producing residual connection $Res_i$ and skip connection $Skip_i$ at each block, where residual outputs are propagated to subsequent blocks and skip outputs are aggregated to form the final prediction of score $s_{\theta}$. The input $\tilde{x}_{t+1:t+\tau}$ denotes the noised LOB trajectory at a given diffusion step. $Dim_{emb}$ is price level embedding, which enables the model to distinguish between spatially different price levels in the order book.

\paragraph{Regime Condition Encoders.}
Future market regimes are encoded through a condition encoder (Figure~\ref{fig:architecture}(b)).
We distinguish between $Local$ and $Global$ regime information. Local conditions capture step-wise signals, including historical LOB, liquidity, imbalance, and time-of-day information, and are encoded using convolutional layers. Global conditions represent longer-horizon market characteristics, such as future trend and volatility, and are encoded using multilayer perceptrons.
This separation allows DiffLOB to model heterogeneous regime effects at different temporal scales.

\paragraph{Control Module and Dual-stage Training.}
To enforce controlling over generated trajectories, DiffLOB introduces a control module inspired by ControlNet ~\citep{zhang2023adding}, as shown in (Figure~\ref{fig:architecture}(c)).
The condition encoder and wavenet module are first trained to model the base data distribution. Subsequently, the control module is trained while keeping the other backbone parameters frozen. Control signals are injected additively through zero-initialized $1\times1$ convolutional layers, whose weights and biases are initialized to zero. As a result, the control pathway has no effect at initialization and gradually learns regime-specific interventions in a stable and interpretable manner.

\paragraph{Wavenet Module with FiLM Modulation.}
Each residual block follows a Wavenet-style gated architecture with residual and skip connections, as shown in Figure~\ref{fig:architecture}(d).
Conditioning information is injected using feature-wise linear modulation (FiLM) ~\citep{perez2018film}.
Block activations are sequentially modulated by the diffusion timestep embedding $t_{emb}$, $Local$ and $Global$ regime embeddings.
This sequential modulation scheme enables structured interaction between diffusion time, short-term market states, and long-horizon regime characteristics.

\section{Experimental Setup}

In this section, we provide data, preprocessing methods, as well as the training and sampling procedures. 

\subsection{Data and Preprocessing}

We use the LOBSTER data\footnote{https://lobsterdata.com/} as our LOB data source ~\citep{huang2011lobster}. We sample one snapshot per second. Each snapshot includes the top 10 levels on both the bid and ask sides. In particular, we train, evaluate, and test our model separately on three stocks, which are AMZN, AAPL and GOOG. For each stock, we use 16 consecutive trading days for training (1 February to 23 February 2023), 1 day for validation (24 February  2023), and the final 2 days for testing (27-28 February 2023).



\paragraph{Price Representation.}
Let $a_t^{(k)}$ and $b_t^{(k)}$ denote the ask and bid prices at level $k$ at time $t$.
We define the mid-price as
\begin{equation*}
m_t = \frac{a_t^{(1)} + b_t^{(1)}}{2}.
\end{equation*}
To model temporal price dynamics, we use one-step mid-price difference:
\begin{equation*}
\label{eq:mid-price-difference}
r_t = m_{t+1} - m_t.
\end{equation*}
In addition, we include cross-sectional price differences to capture the spatial structure of the order book:
\begin{equation*}
\label{eq:cross-sectional price difference}
\resizebox{.91\linewidth}{!}{$
\Delta a_t^{(k)} = a_t^{(k)} - a_t^{(k-1)}, \quad
\Delta s_t = a_t^{(1)} - b_t^{(1)}, \quad
\Delta b_t^{(k)} = b_t^{(k-1)} - b_t^{(k)}
$}
\end{equation*}
for $k=2,\dots,K$. The final price representation concatenates the one-step mid-price difference and the cross-sectional price differences.
This construction preserves the original dimension while removing absolute price levels, resulting in a more stable and learnable representation for generative modeling.

\paragraph{Volume Representation.}
Raw volume $v$ exhibit heavy-tailed distributions and large scale variations.
To stabilize training, we cap volume values at the 99th percentile followed by a square-root normalization:
\begin{equation*}
\tilde{v} = \frac{\sqrt{v}}{c},
\end{equation*}
where $c$ is a constant scaling factor.

\paragraph{Market Regimes.}

Market regimes are computed from the target LOB trajectory $x_{t+1:t+\tau}$ and used as conditioning variables.
Specifically, we define:
\begin{itemize}
    \item \textbf{Trend} as the cumulative mid-price return over $\tau$:
    \begin{equation*}
    c^{\text{trend}}_{t+1:t+\tau} = \sum_{i=0}^{\tau-1}r(t+i).
    \end{equation*}
    \item \textbf{Volatility} as the standard deviation of returns over $\tau$:
    \begin{equation*}
    c^{\text{vol}}_{t+1:t+\tau} = \sqrt{\frac{1}{\tau}\sum_{i=0}^{\tau-1}r_{t+i}^2-\left(\frac{1}{\tau}\sum_{i=0}^{\tau-1}r_{t+i}\right)^2}.
    \end{equation*}
    \item \textbf{Liquidity} as the total standing volume over each timestep:
    \begin{equation*}
    c^{\text{liq}}_{t+i} = \sum_{k=1}^{K} \left(v_{t+i}^{a,(k)} + v_{t+i}^{b,(k)}\right).
    \end{equation*}
    \item \textbf{Order-flow imbalance} as the normalized volume imbalance over each timestep:
    \begin{equation*}
    c^{\text{imb}}_{t+i} =
    \frac{\sum_{k=1}^{K} \left(v_{t+i}^{a,(k)} - v_{t+i}^{b,(k)}\right)}
    {\sum_{k=1}^{K} \left(v_{t+i}^{a,(k)} + v_{t+i}^{b,(k)}\right)}.
    \end{equation*}
\end{itemize}

\subsection{Training and Sampling}

Historical length $t$ and generated length $\tau$ are fixed to 32 for all experiments. To promote stable training and reduce overfitting, we employ early stopping and exponential moving average (EMA) of model parameters.
Early stopping is based on the validation loss and terminates training if no improvement greater than $0.001$ is observed over $100$ epochs.
The model is optimized using Adam~\citep{diederik2014adam} with a learning rate of $1\times10^{-4}$ and a batch size of $128$. We adopt the DDPM parameterization with $100$ discrete noise levels and generate samples via ancestral sampling~\citep{ho2020denoising}.
Conditions are incorporated during training and sampling using classifier-free guidance~\citep{ho2022classifier}.
The network consists of $16$ residual blocks with SiLU activations applied after each convolution. 
Complete training and sampling algorithm are provided in Appendix \ref{sec:algo}.

\section{Experimental Results}

We evaluate DiffLOB from three complementary perspectives to assess its controllable realism, counterfactual validity, and practical usefulness. We compare DiffLOB with several baseline generative models. These include diffusion-based baselines (Diff-CSDI ~\citep{tashiro2021csdi} and Diff-S4~\citep{alcaraz2022diffusion}), non-diffusion generative models (cGAN~\citep{cont2023limit} and cVAE~\citep{sohn2015learning}), and an autoregressive S4-based model AR~\citep{nagy2023generative}. DiffLOB without the control module is also included for ablation study use. For clarity of presentation, all figures use AMZN as an illustrative example. 

\subsection{Controllable Realism}

\begin{table*}[!htbp]
\centering
\scalebox{0.72}{
\begin{tabular}{@{}lrrrrrrrrrrrr@{}}
\toprule
               & \multicolumn{4}{c}{AMZN}                                                                                   & \multicolumn{4}{c}{AAPL}                                                                                   & \multicolumn{4}{c}{GOOG}                                                                                   \\ \midrule
Price-Realism  & \multicolumn{1}{c}{KS} & \multicolumn{1}{c}{Wasserstein} & \multicolumn{1}{c}{KL} & \multicolumn{1}{c}{JS} & \multicolumn{1}{c}{KS} & \multicolumn{1}{c}{Wasserstein} & \multicolumn{1}{c}{KL} & \multicolumn{1}{c}{JS} & \multicolumn{1}{c}{KS} & \multicolumn{1}{c}{Wasserstein} & \multicolumn{1}{c}{KL} & \multicolumn{1}{c}{JS} \\ \midrule
DiffLOB        & \textbf{0.052384}      & \textbf{0.028371}               & \textbf{0.075605}      & \textbf{0.017585}      & 0.031695               & \textbf{0.016446}               & \textbf{0.026561}      & \textbf{0.005704}      & \textbf{0.055462}      & \textbf{0.035195}               & 0.363414               & 0.037249               \\
DiffLOB w/o C   & 0.212398               & 0.182775                        & 0.730008               & 0.064836               & 0.079402               & 0.064358                        & 0.095074               & 0.014745               & 0.103601               & 0.085803                        & \textbf{0.30239}       & \textbf{0.027799}      \\
Diff-CSDI      & 0.628793               & 64.621056                       & 5.862505               & 0.306881               & 0.576959               & 68.690042                       & 5.507347               & 0.286949               & 0.573082               & 47.207406                       & 4.99384                & 0.261333               \\
Diff-S4        & 0.200429               & 0.155586                        & 0.304272               & 0.051942               & 0.101795               & 0.0915                          & 0.149292               & 0.019214               & 0.153828               & 0.06162                         & 0.603084               & 0.058418               \\
CGAN           & 0.053521               & 0.048909                        & 0.378863               & 0.026156               & 0.830413               & 2.402774                        & 9.343481               & 0.466478               & 0.28967                & 0.296793                        & 1.523045               & 0.114271               \\
CVAE           & 0.173396               & 0.142745                        & 0.406215               & 0.053637               & \textbf{0.025862}      & 0.021286                        & 0.135777               & 0.013759               & 0.155807               & 0.085175                        & 0.419525               & 0.050767               \\
AR             & 0.118031               & 0.076163                        & 0.419127               & 0.037131               & 0.173451               & 0.145696                        & 0.304234               & 0.03807                & 0.107334               & 0.046178                        & 0.324453               & 0.040902               \\ \midrule
Volume-Realism & \multicolumn{1}{c}{KS} & \multicolumn{1}{c}{Wasserstein} & \multicolumn{1}{c}{KL} & \multicolumn{1}{c}{JS} & \multicolumn{1}{c}{KS} & \multicolumn{1}{c}{Wasserstein} & \multicolumn{1}{c}{KL} & \multicolumn{1}{c}{JS} & \multicolumn{1}{c}{KS} & \multicolumn{1}{c}{Wasserstein} & \multicolumn{1}{c}{KL} & \multicolumn{1}{c}{JS} \\ \midrule
DiffLOB        & \textbf{0.109174}      & \textbf{97.54283}               & 0.111954               & 0.023297               & \textbf{0.087666}      & \textbf{55.377778}              & \textbf{0.071645}      & \textbf{0.014677}      & \textbf{0.099899}      & \textbf{97.270339}              & \textbf{0.086374}      & \textbf{0.016609}      \\
DiffLOB w/o C   & 0.113053               & 117.36868                       & \textbf{0.108419}      & \textbf{0.022918}      & 0.100573               & 72.929232                       & 0.091735               & 0.019714               & 0.126238               & 132.065501                      & 0.117649               & 0.023473               \\
Diff-CSDI      & 0.293449               & 599243.7598                     & 0.324422               & 0.052769               & 0.235838               & 587451.9702                     & 0.29902                & 0.04794                & 0.256603               & 578472.3812                     & 0.266948               & 0.039498               \\
Diff-S4        & 0.170321               & 161.665895                      & 0.178012               & 0.036454               & 0.132301               & 104.168332                      & 0.138386               & 0.025026               & 0.246833               & 279.02256                       & 0.325909               & 0.060015               \\
CGAN           & 0.288535               & 321.049535                      & 0.500225               & 0.090229               & 0.487548               & 275.698425                      & 2.956931               & 0.23594                & 0.127321               & 151.897776                      & 0.137742               & 0.027589               \\
CVAE           & 0.260795               & 255.043006                      & 1.154918               & 0.153781               & 0.23728                & 169.905872                      & 1.251494               & 0.140914               & 0.231203               & 240.882982                      & 0.864015               & 0.112463               \\
AR             & 0.397229               & 444.216686                      & 1.172602               & 0.159319               & 0.285415               & 188.199251                      & 0.669676               & 0.112751               & 0.2863                 & 343.938149                      & 0.724715               & 0.106982               \\ \bottomrule
\end{tabular}}
\caption{Controllable Realism on Three Stocks.}
\label{tab:realism}
\end{table*}

We first evaluate the realism of DiffLOB by conditioning the model on \emph{observed} future market regimes and comparing the generated trajectories with real data.
The goal of this evaluation is to assess whether DiffLOB can faithfully reproduce key statistics of LOB dynamics.

\begin{figure}[!htbp]
  \centering
  
  \begin{subfigure}[t]{\linewidth}
    \centering
    \includegraphics[width=\linewidth]{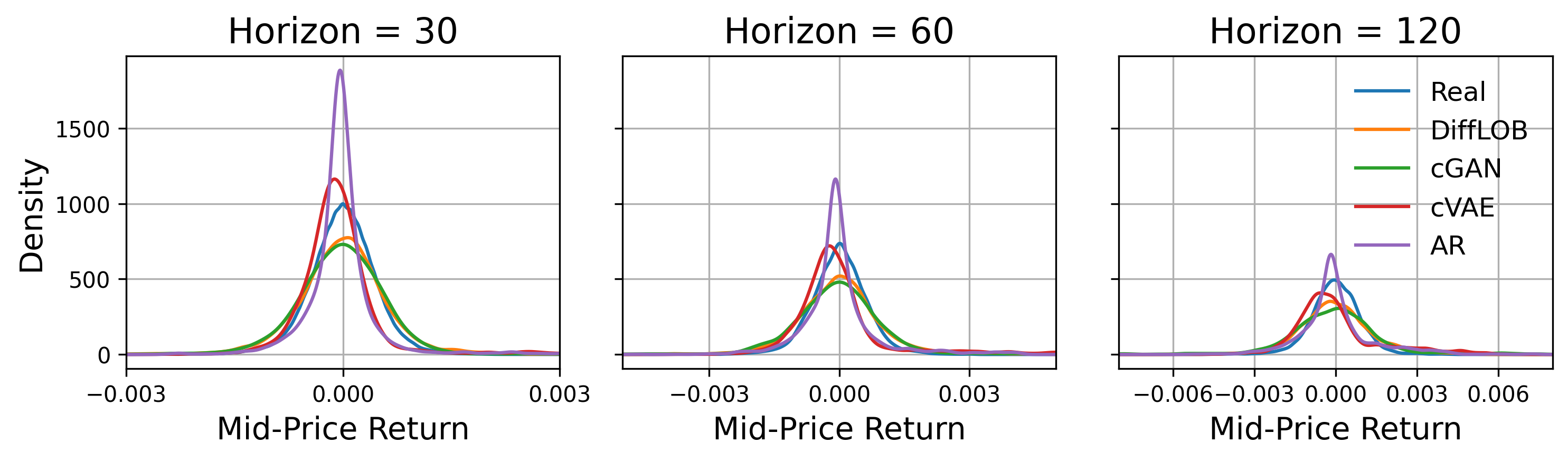}
    \caption{Mid Price Return on Different Horizons.}
    \label{fig:mid-price-return}
  \end{subfigure}

   \begin{subfigure}[t]{\linewidth}
    \centering
    \includegraphics[width=\linewidth]{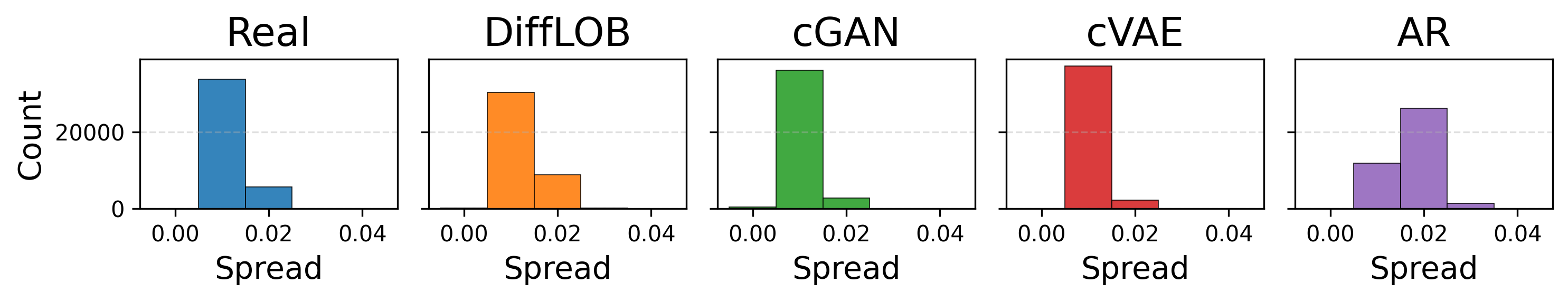}
    \caption{Spread Distribution.}
    \label{fig:spread}
  \end{subfigure}

  \begin{subfigure}[t]{\linewidth}
    \centering
    \includegraphics[width=\linewidth]{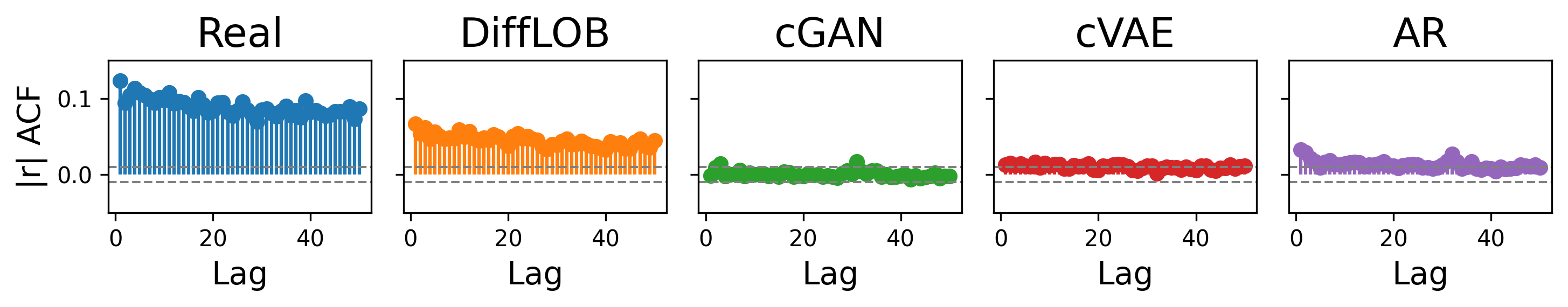}
    \caption{Volatility Clustering.}
    \label{fig:acf}
  \end{subfigure}

  \caption{Realism on Price.}
  \label{fig:price-realism}
\end{figure}

Figure~\ref{fig:price-realism} evaluates the realism of price dynamics generated by models. Figure~\ref{fig:mid-price-return} shows that DiffLOB closely matches the empirical distributions of mid-price returns across multiple horizons. Figure~\ref{fig:spread} reports the distribution of bid--ask spread, where DiffLOB replicates real distribution best. Figure~\ref{fig:acf} compares the autocorrelation of absolute returns. DiffLOB reproduces the persistent decay pattern characteristic of volatility clustering, whereas cGAN and cVAE largely fail to capture temporal dependence and AR models underestimate long-range correlations. 

\begin{figure*}[!htbp]  
    \centering  
    \includegraphics[width=\textwidth]{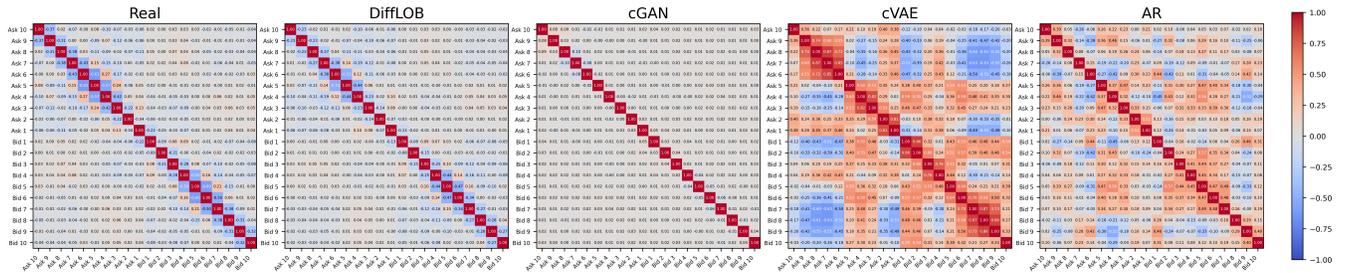}  
    \caption{Temporal Difference Volume Correlation.}
    \label{fig:volume-correlation}  
\end{figure*}

Figure \ref{fig:volume-correlation} is calculated from the first-order differences of the volume snapshots ($v_{t+1} - v_{t}$). A notable characteristic in the temporal difference correlation structure is the negative correlation at adjacent price levels, which is captured only by DiffLOB. We also show more details about marginal volume distribution across price levels in Appendix \ref{sec:volume-distribution}.

Figure~\ref{fig:real-condition-distribution} visualizes the distributions of future market regime variables under observed conditions, including trend, volatility, liquidity, and order-flow imbalance. DiffLOB accurately captures the central tendency of trend and volatility, while achieving an excellent match to the full empirical distributions of liquidity and order-flow imbalance, including both spread and tail behavior. In contrast, baseline models exhibit noticeable mismatches, such as overly concentrated distributions (AR on Trend) and shifted modes (cVAE on Liquidity, AR on Volatility, AR on Imbalance). These results further confirm that DiffLOB can faithfully reproduce regime-level statistics when conditioned on true future regimes, supporting the quantitative findings in Table~\ref{tab:realism}.

\begin{figure}[t]  
    \centering  
    \includegraphics[width=\linewidth]{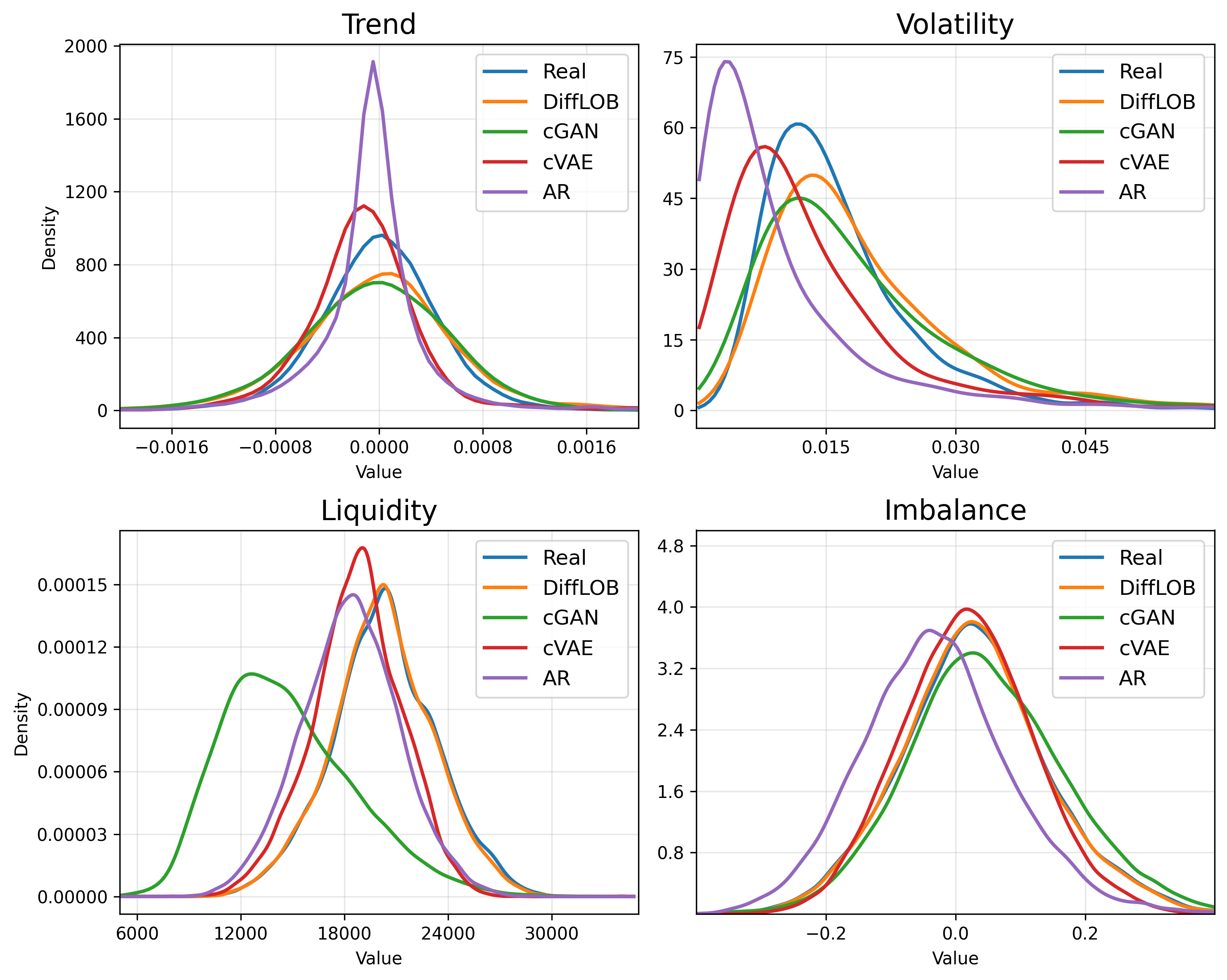}  
    \caption{Controllable Realism Distribution..}
    \label{fig:real-condition-distribution}  
\end{figure}

Table~\ref{tab:realism} reports quantitative distributional distances between generated and real LOB trajectories under observed future regimes on the three stocks across both price-related and volume-related statistics. 
We consider four distance metrics, including Kolmogorov–Smirnov (KS) statistic, Wasserstein distance, Kullback–Leibler (KL) divergence, and Jensen–Shannon (JS) divergence. Across all assets and metrics, DiffLOB consistently achieves the lowest or near-lowest distances, indicating that it best matches the empirical distributions of real data when conditioned on true future regimes. Removing the control module (DiffLOB w/o C) leads to a clear degradation in performance, highlighting the importance of explicit regime control even under observed conditions. Diffusion-based baselines (Diff-CSDI and Diff-S4) improve upon non-diffusion methods in some cases, but remain substantially less accurate than DiffLOB, showing the effectiveness of our proposed DiffLOB architecture described in Section ~\ref{subsec:difflob-architecture}.
Non-diffusion baselines, including cGAN, cVAE, and the autoregressive model, exhibit significantly larger discrepancies, especially in Wasserstein and KL distances. Overall, these results demonstrate that DiffLOB achieves superior controllable realism across both price and volume dimensions, and that designed control module plays a critical role in accurately reproducing LOB statistics.

\subsection{Counterfactual Validity}

We next evaluate the counterfactual validity by intervening on future market regimes and examining the resulting generated trajectories. Specifically, we impose extreme hypothetical conditions, including high and low trend, volatility, liquidity, and order-flow imbalance, and assess whether the generated samples exhibit statistical properties consistent with the imposed regimes.

\begin{figure}[!htbp]  
    \centering  
    \includegraphics[width=\linewidth]{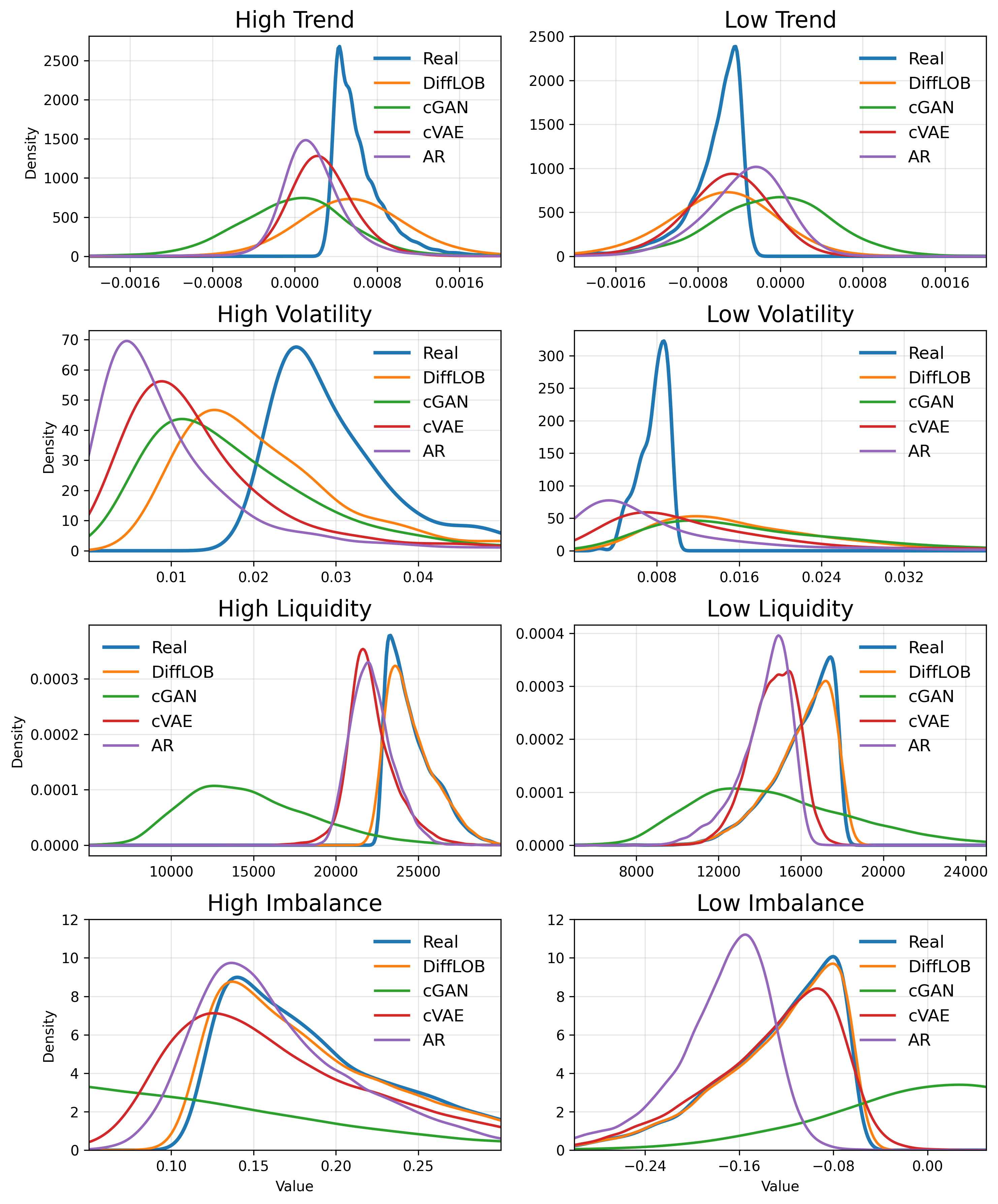}  
    \caption{Counterfactual Realism Distribution.}
    \label{fig:counterfactual-distribution}  
\end{figure}

Figure~\ref{fig:counterfactual-distribution} evaluates counterfactual validity under explicit regime interventions.
For each regime variable, we compare distributions generated under high and low counterfactual conditions with real trajectories in the corresponding extreme regimes (top and bottom 20\%).
For trend and volatility, DiffLOB does not perfectly overlap with empirical distributions but consistently shifts in the correct direction and preserves the separation between high and low conditions.
In contrast, for liquidity and order-flow imbalance, DiffLOB closely matches the empirical distributions, indicating strong controllability for volume-driven dynamics.
Overall, these results show that DiffLOB learns an interpretable and intervention-consistent relationship between future regimes and LOB dynamics, enabling meaningful counterfactual generation.

\begin{figure}[!htbp]  
    \centering  
    \includegraphics[width=\linewidth]{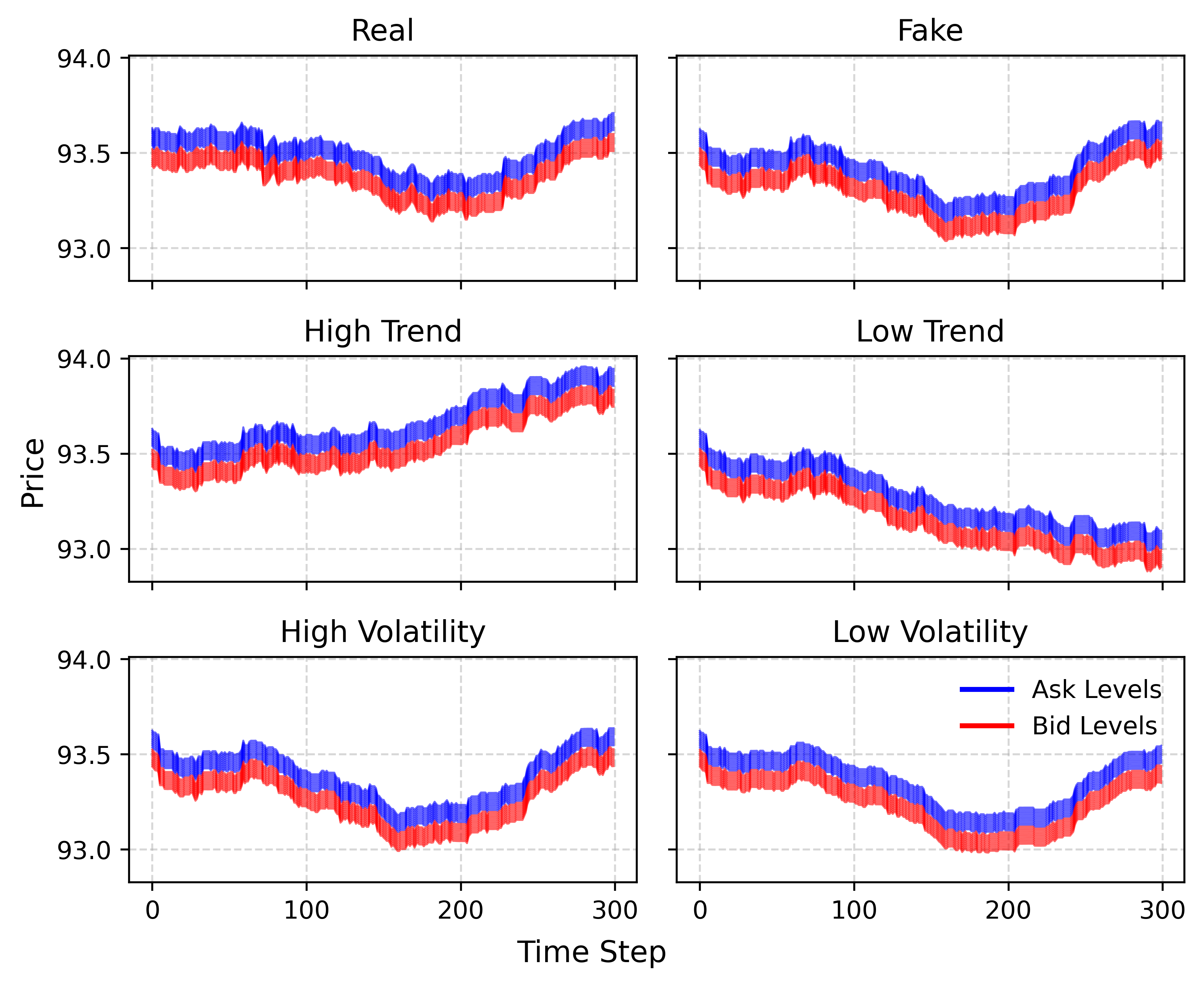}  
    \caption{Counterfactual LOB Price Trajectories.}
    \label{fig:counterfactual-price}  
\end{figure}

\begin{figure}[!htbp]  
    \centering  
    \includegraphics[width=\linewidth]{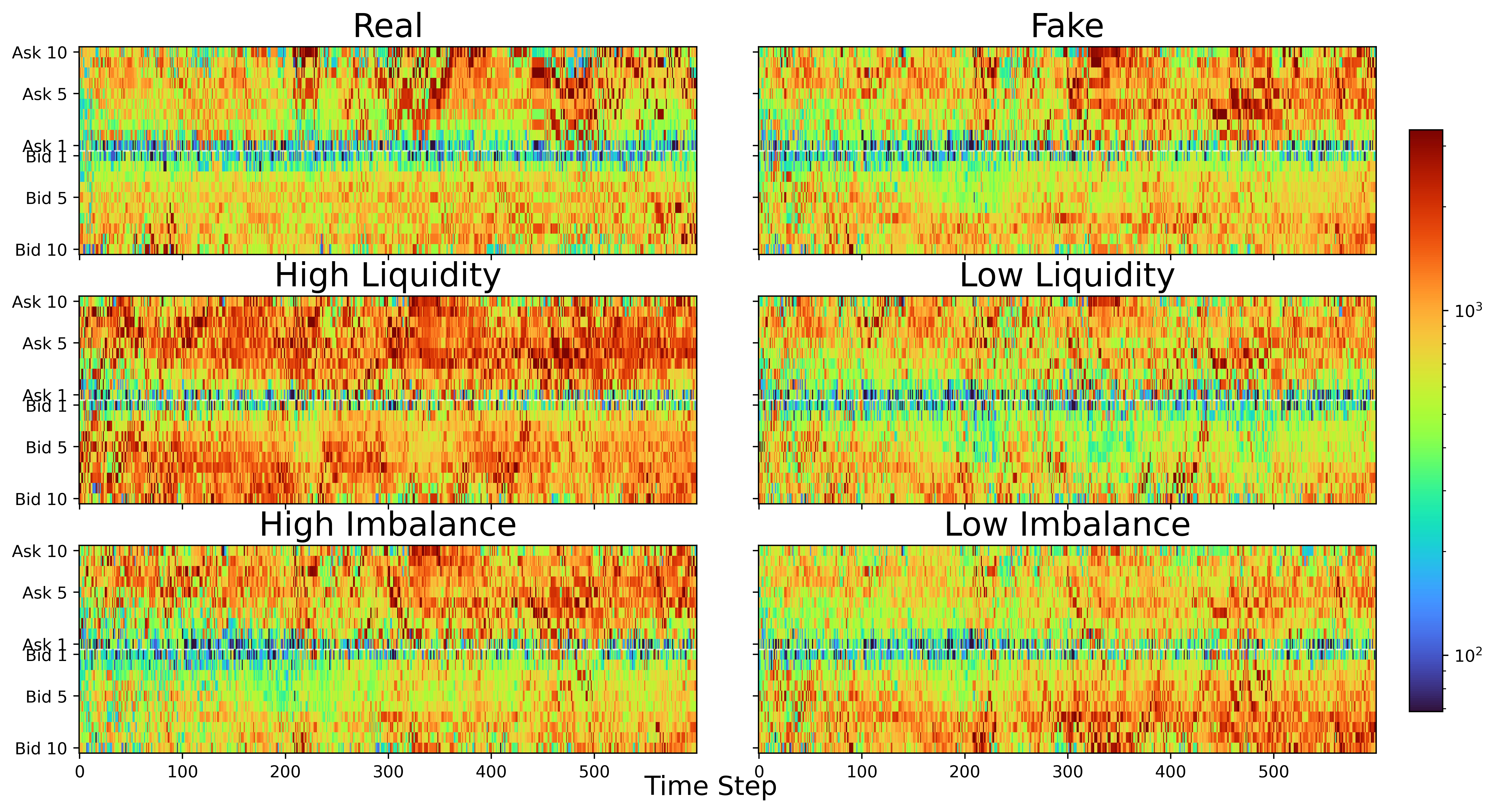}  
    \caption{Counterfactual LOB Volume Trajectories.}
    \label{fig:counterfactual-volume}  
\end{figure}

Figure~\ref{fig:counterfactual-price} and Figure~\ref{fig:counterfactual-volume} provide  visualizations of counterfactual generation on price and volume trajectories.
In Figure~\ref{fig:counterfactual-price}, explicit interventions on future trend and volatility induce coherent and interpretable changes.
High-trend conditions lead to persistent upward price movements, while low-trend conditions result in declining trajectories.
Similarly, high-volatility interventions produce more volatile price paths with larger fluctuations, whereas low-volatility conditions yield smoother dynamics.
In Figure~\ref{fig:counterfactual-volume}, imposing high-liquidity conditions leads to consistently larger volume magnitudes across the LOB, while low-liquidity conditions shift the overall volume downward.
Imbalance interventions induce clear asymmetries between bid and ask sides, consistent with the imposed order-flow imbalance.
These results demonstrate that DiffLOB can generate regime-consistent dynamics beyond matching marginal distributions.

\begin{table}[!htbp]
\centering
\scalebox{0.8}{
\begin{tabular}{@{}lrrrr@{}}
\toprule
High-Trend     & \multicolumn{1}{c}{KS} & \multicolumn{1}{c}{Wasserstein} & \multicolumn{1}{c}{KL} & \multicolumn{1}{c}{JS} \\ \midrule
DiffLOB        & 0.444095               & 0.522512                        & 2.517532               & 0.180279               \\
DiffLOB w/o C   & 0.515225               & 0.801835                        & 4.879508               & 0.265181               \\
Diff-CSDI      & 0.627345               & 65.274135                       & 5.855148               & 0.307245               \\
Diff-S4        & 0.292354               & 0.264849                        & 1.098681               & 0.092509               \\
CGAN           & \textbf{0.048718}      & \textbf{0.036091}               & \textbf{0.422542}      & \textbf{0.026361}      \\
CVAE           & 0.298339               & 0.225797                        & 0.79771                & 0.092942               \\
AR             & 0.200438               & 0.199092                        & 1.210306               & 0.091039               \\ \midrule
Low-Trend      & \multicolumn{1}{c}{KS} & \multicolumn{1}{c}{Wasserstein} & \multicolumn{1}{c}{KL} & \multicolumn{1}{c}{JS} \\ \midrule
DiffLOB        & 0.425788               & 0.507738                        & 3.529521               & 0.182637               \\
DiffLOB w/o C   & 0.339051               & 0.36766                         & 2.525023               & 0.141955               \\
Diff-CSDI      & 0.755996               & 84.376958                       & 6.563037               & 0.356456               \\
Diff-S4        & 0.431509               & 0.521119                        & 3.855445               & 0.187511               \\
CGAN           & \textbf{0.039677}      & \textbf{0.044298}               & \textbf{0.47366}       & \textbf{0.023752}      \\
CVAE           & 0.38066                & 0.459526                        & 2.89087                & 0.156796               \\
AR             & 0.264977               & 0.296113                        & 1.946188               & 0.107689               \\ \midrule
High-Volatility & \multicolumn{1}{c}{KS} & \multicolumn{1}{c}{Wasserstein} & \multicolumn{1}{c}{KL} & \multicolumn{1}{c}{JS} \\ \midrule
DiffLOB        & \textbf{0.083474}      & \textbf{0.06075}                & \textbf{0.169118}      & \textbf{0.03113}       \\
DiffLOB w/o C   & 0.292613               & 0.237796                        & 1.004272               & 0.108261               \\
Diff-CSDI      & 0.609586               & 59.726185                       & 5.559404               & 0.286991               \\
Diff-S4        & 0.118421               & 0.12436                         & 0.67194                & 0.064088               \\
CGAN           & 0.178244               & 0.088973                        & 0.524627               & 0.048186               \\
CVAE           & 0.118761               & 0.077975                        & 0.646276               & 0.071165               \\
AR             & 0.083668               & 0.065337                        & 0.454682               & 0.041336               \\ \midrule
Low-Volatility & \multicolumn{1}{c}{KS} & \multicolumn{1}{c}{Wasserstein} & \multicolumn{1}{c}{KL} & \multicolumn{1}{c}{JS} \\ \midrule
DiffLOB        & \textbf{0.150047}      & \textbf{0.072569}               & \textbf{0.772114}      & \textbf{0.061085}      \\
DiffLOB w/o C   & 0.163749               & 0.140627                        & 1.38697                & 0.075512               \\
Diff-CSDI      & 0.636904               & 59.84074                        & 5.663091               & 0.293081               \\
Diff-S4        & 0.297515               & 0.160111                        & 0.999069               & 0.111148               \\
CGAN           & 0.16065                & 0.113201                        & 1.285274               & 0.078348               \\
CVAE           & 0.287089               & 0.16455                         & 1.092421               & 0.100044               \\
AR             & 0.244042               & 0.141                           & 1.457307               & 0.09852                \\ \midrule
High-Liquidity & \multicolumn{1}{c}{KS} & \multicolumn{1}{c}{Wasserstein} & \multicolumn{1}{c}{KL} & \multicolumn{1}{c}{JS} \\ \midrule
DiffLOB        & 0.137538               & \textbf{168.024319}             & 0.208135               & 0.034988               \\
DiffLOB w/o C   & \textbf{0.134555}      & 179.447363                      & \textbf{0.18559}       & \textbf{0.03162}       \\
Diff-CSDI      & 0.318472               & 617445.7056                     & 0.419245               & 0.05283                \\
Diff-S4        & 0.199552               & 229.470802                      & 0.279902               & 0.050367               \\
CGAN           & 0.389481               & 513.984844                      & 1.1278                 & 0.153755               \\
CVAE           & 0.335492               & 365.126803                      & 1.982426               & 0.229481               \\
AR             & 0.422048               & 445.418425                      & 1.840147               & 0.215084               \\ \midrule
Low-Liquidity  & \multicolumn{1}{c}{KS} & \multicolumn{1}{c}{Wasserstein} & \multicolumn{1}{c}{KL} & \multicolumn{1}{c}{JS} \\ \midrule
DiffLOB        & \textbf{0.126894}      & \textbf{104.19346}              & \textbf{0.099957}      & \textbf{0.021029}      \\
DiffLOB w/o C   & 0.140489               & 110.790344                      & 0.114877               & 0.025278               \\
Diff-CSDI      & 0.311335               & 585813.0602                     & 0.444758               & 0.073959               \\
Diff-S4        & 0.224048               & 142.265977                      & 0.212572               & 0.048326               \\
CGAN           & 0.214546               & 167.358666                      & 0.315572               & 0.063397               \\
CVAE           & 0.351886
               & 231.117557
                      & 1.748266
               & 0.220437
                \\
AR             & 0.410027               & 274.375898                      & 1.572332               & 0.19979                \\ \midrule
High-Imbalance & \multicolumn{1}{c}{KS} & \multicolumn{1}{c}{Wasserstein} & \multicolumn{1}{c}{KL} & \multicolumn{1}{c}{JS} \\ \midrule
DiffLOB        & 0.142787               & \textbf{154.156402}             & 0.182522               & 0.032153               \\
DiffLOB w/o C   & \textbf{0.130096}      & 155.84261                       & 0.181721      & \textbf{0.032042}      \\
Diff-CSDI      & 0.303622               & 629928.7738                     & 0.363013               & 0.054523               \\
Diff-S4        & 0.163933               & 165.329592                      & \textbf{0.180472}              & 0.035623               \\
CGAN           & 0.285831               & 342.915205                      & 0.59001                & 0.095159               \\
CVAE           & 0.274923               & 291.958093                      & 1.485293               & 0.171144               \\
AR             & 0.432061               & 397.430331                      & 1.656605               & 0.202636               \\ \midrule
Low-Imbalance  & \multicolumn{1}{c}{KS} & \multicolumn{1}{c}{Wasserstein} & \multicolumn{1}{c}{KL} & \multicolumn{1}{c}{JS} \\ \midrule
DiffLOB        & \textbf{0.138717}      & \textbf{147.853412}             & \textbf{0.181154}      & \textbf{0.034248}      \\
DiffLOB w/o C   & 0.153997               & 174.322966                      & 0.204804               & 0.038629               \\
Diff-CSDI      & 0.307549               & 593616.0608                     & 0.43386                & 0.063705               \\
Diff-S4        & 0.198524               & 209.414158                      & 0.253176               & 0.049452               \\
CGAN           & 0.298427               & 356.49669                       & 0.635989               & 0.101998               \\
CVAE           & 0.290173               & 290.269725                      & 1.529841               & 0.18089                \\
AR             & 0.383488               & 338.46483                       & 1.580578               & 0.186188               \\ \bottomrule
\end{tabular}}
\caption{Counterfactual Validity on AMZN.}
\label{tab:counterfactual-validity}
\end{table}

Table~\ref{tab:counterfactual-validity} reports quantitative distances computed between counterfactual samples generated with a fixed future regime (e.g., high trend) and real market trajectories whose regime values fall into the same extreme quantile (top or bottom 20\%).
cGAN performs competitively on trend-related metrics but degrades substantially for volatility, liquidity, and imbalance, indicating limited regime robustness.
DiffLOB without the control module performs relatively well under high-liquidity and high-imbalance conditions, suggesting that volume-driven regimes can be partially captured by the diffusion backbone alone, consistent with findings in \citep{wang2025diffvolume}.
However, across the majority of regimes and distance metrics, DiffLOB consistently achieves the lowest or near-lowest distances.
This demonstrates that explicitly modeling future regimes through the control module leads to more reliable and coherent counterfactual responses across diverse market conditions. Please see the complete table on the three stocks in Appendix ~\ref{sec:complete-counterfactual-table}.

\subsection{Counterfactual Usefulness}

Finally, we assess the practical usefulness of counterfactual trajectories generated by DiffLOB. 
We consider two tasks: trend prediction and liquidity prediction.
Trend prediction is formulated as a classification task that predicts the direction of future price movement, and is evaluated using accuracy ($Acc$). Liquidity prediction is treated as a regression task that predicts future liquidity values, evaluated using the coefficient of determination ($R^2$).
For both tasks, models are trained on past 1 minute LOB data to predict future 1 minute regime. Models are evaluated on 1 March 2023, with performance reported separately on top 20\% and bottom 20\% of the regime distribution.
We compare three training settings: \texttt{Real}, using only real samples; \texttt{Real * 2}, duplicating real data to control for dataset size; and \texttt{Real+CF}, augmenting real data with counterfactual trajectories to enrich extreme regime coverage.

\begin{table}[!htbp]
\centering
\scalebox{0.7}{
\begin{tabular}{@{}lrrrrrr@{}}
\toprule
                     & \multicolumn{2}{c}{AMZN}                                                                     & \multicolumn{2}{c}{AAPL}                                                                     & \multicolumn{2}{c}{GOOG}                                                                     \\ \midrule
Trend \\ Prediction     & \multicolumn{1}{c}{Acc-High}                  & \multicolumn{1}{c}{Acc-Low}                  & \multicolumn{1}{c}{Acc-High}                  & \multicolumn{1}{c}{Acc-Low}                  & \multicolumn{1}{c}{Acc-High}                  & \multicolumn{1}{c}{Acc-Low}                  \\ \midrule
Real                 & \textbf{0.983}                                & 0.061                                        & 0.885                                         & 0.14                                         & 0.517                                         & 0.453                                        \\
Real * 2             & 0.98                                          & 0.064                                        & 0.89                                          & 0.153                                        & 0.5                                           & 0.488                                        \\
Real + CF            & 0.944                                         & \textbf{0.289}                               & \textbf{0.918}                                & \textbf{0.17}                                & \textbf{0.56}                                 & \textbf{0.556}                               \\ \midrule
Liquidity \\ Prediction & \multicolumn{1}{c}{R\textasciicircum{}2-High} & \multicolumn{1}{c}{R\textasciicircum{}2-Low} & \multicolumn{1}{c}{R\textasciicircum{}2-High} & \multicolumn{1}{c}{R\textasciicircum{}2-Low} & \multicolumn{1}{c}{R\textasciicircum{}2-High} & \multicolumn{1}{c}{R\textasciicircum{}2-Low} \\ \midrule
Real                 & -1.568                                        & 0.086                                        & -2.512                                        & -0.184                                       & -0.742                                        & -0.039                                       \\
Real * 2             & -1.656                                        & 0.127                                        & -2.525                                        & -0.26                                        & -0.796                                        & \textbf{-0.011}                              \\
Real + CF            & \textbf{-1.509}                               & \textbf{0.278}                               & \textbf{-1.432}                               & \textbf{0.17}                                & \textbf{0.017}                                & -0.053                                       \\ \bottomrule
\end{tabular}}
\caption{Counterfactual Usefulness on the three Stocks.}
\label{tab:usefulness}
\end{table}

In Table~\ref{tab:usefulness}, we can see that augmenting real data with counterfactual samples improves performance on most high- and low-regime subsets, compared to training on real data alone.
The gains demonstrate that counterfactual trajectories provide complementary information that is scarce in real data, proving the usefulness of counterfactual trajectories generated by DiffLOB.

\section{Conclusion}

We introduce DiffLOB, a diffusion-based framework for controllable and counterfactual generation of limit order book trajectories.
By explicitly conditioning on future market regimes—including trend, volatility, liquidity, and order-flow imbalance—DiffLOB enables direct intervention on hypothetical future conditions.
Extensive experiments demonstrate that DiffLOB achieves superior controllable realism, generates coherent and regime-consistent counterfactual trajectories, and provides tangible benefits for downstream prediction tasks under extreme market regimes.
These results highlight the importance of explicit regime-aware control for realistic simulation and meaningful counterfactual analysis.

\clearpage
\bibliographystyle{named}
\bibliography{ijcai26}

\clearpage         
\appendix          

\section{Diffusion Models Theory}
\label{sec:theory}
In the appendix, the variable $t$ denotes the \emph{diffusion time step} (or noise level) used in the diffusion process, rather than the real-world market time index.
This is distinct from the notation $c_{t+1:t+\tau}$ in the main text, where $t$ refers to the current market time and $t+1:t+\tau$ denotes a future horizon.
The two uses of $t$ should not be confused.

\smallskip 
\noindent \textbf{DDPMs and Stochastic Differential Equations.} \citet{song2020score} demonstrate that DDPMs can be understood from the perspective of stochastic differential equations (SDEs). Let $\{\mathbf{x}(t) \}_{t=0}^T$ be a stochastic diffusion process indexed by a continuous time variable $t\in[0, T]$, evolving from $\mathbf{x}(0) \sim p_0$, the true data distribution, to $\mathbf{x}(T) \sim p_T$, approximately the tractable prior distribution. Denote the probability density function of $\mathbf{x}(t)$ by $p_t(\mathbf{x})$ and the transition kernel from $\mathbf{x}(s)$ to $\mathbf{x}(t)$ by $p_{st}(\mathbf{x}(t)|\mathbf{x}(s))$, for $0\leq s<t \leq T$. Then, we can use an SDE to represent such a forward diffusion process: 
\begin{equation}
    d\mathbf{x}=\mathbf{f}(\mathbf{x},t)\ dt + g(t) \ d \mathbf{w},
\label{eq:sde-forward-process}
\end{equation}
where $\mathbf{f}(\mathbf{x},t) dt$ is referred to as the \textit{drift} term, and $ g(t) d \mathbf{w}$ is referred to as the \textit{diffusion} term. Here, $\mathbf{w}$ is a standard Wiener process and $d\mathbf{w} \sim \mathcal{N}(0, dt \mathbf{I})$. The synthetic data generation process is the reverse process of Eq. (\ref{eq:sde-forward-process}), which is also an SDE~\cite{anderson1982reverse}:
\begin{equation}
    d\mathbf{x}=[\mathbf{f}(\mathbf{x},t)-g^2(t) \nabla_{\mathbf{x}}\log p_t(\mathbf{x})]\ dt +  g(t)\ d \bar{\mathbf{w}},
\label{eq:sde-backward-process}
\end{equation}
where $\bar{\mathbf{w}}$ is a reverse-time Wiener process and  $\nabla_{\mathbf{x}}\log p_t(\mathbf{x})$ is the score of the marginal distribution corresponding to each $t$. It starts from an initial noise sample $\mathbf{x}(T) \sim p_T$ and gradually denoises it step by step following Eq. (\ref{eq:sde-backward-process}). Theoretically, if $T \rightarrow \infty$, we obtain $\mathbf{x}(0) \sim p_0$. To estimate $\nabla_{\mathbf{x}}\log p_t(\mathbf{x})$, the score network $\mathbf{s}_{\mathbf{\theta}}(\mathbf{x}, t)$ is trained using the objective function 
\begin{equation}
\resizebox{0.91\hsize}{!}{$
    \!\!\kappa(t) \mathbb{E}_t  \mathbb{E}_{\mathbf{x}(0)} \mathbb{E}_{\mathbf{x}(t)|\mathbf{x}(0)}\!\big[||\mathbf{s}_{\theta}(\mathbf{x}(t),t)\!- \!\nabla_{\mathbf{x}(t)}\!\log p_{0t}(\mathbf{x}(t)|\mathbf{x}(0))||^2_2 \big],
$}
\label{eq:sde-training}
\end{equation}
where $\kappa:[0,T] \rightarrow \mathbb{R}^+$ is a positive weight and $t \sim \mathcal{U}[0,T]$. Typically, the continuous form of the DDPM forward process is chosen to be
\begin{equation}
    d\mathbf{x}=-\frac{\beta(t)}{2} \mathbf{x}\  dt + \sqrt{\beta(t)} \ d\mathbf{w},
\label{ddpm-continuous-forward-process}
\end{equation}
i.e., $\mathbf{f}(\mathbf{x},t)=-\frac{\beta(t)}{2} \mathbf{x}$ and $g(t) = \sqrt{\beta(t)}$. Substituting $\mathbf{f}(\mathbf{x},t)$ and $g(t)$ in (\ref{eq:sde-backward-process}), we can get the backward process in SDE form for DDPM. 

\smallskip 
\noindent \textbf{Conditional DDPMs.} How do we inject the conditioning $\mathbf{c}$ into the training and sampling process? Here we follow the  \textit{classifier-free guidance} approach~\cite{ho2022classifier}, combining the conditional and unconditional models as follows:
\begin{equation}
    \nabla_{\mathbf{x
    }} \log \tilde{p}(\mathbf{x}|\mathbf{c})=(1+\omega)\nabla_{\mathbf{x}} \log p(\mathbf{x}|\mathbf{c})-\omega\nabla_{\mathbf{x}} \log p(\mathbf{x}),
\label{eq:classifier-free-guidance}
\end{equation}
where $\nabla_{\mathbf{x}} \log p(\mathbf{x}|\mathbf{c})$ represents the conditional  and $\nabla_{\mathbf{x}} \log p(\mathbf{x})$ represents the unconditional score, corresponding to the conditional and unconditional model distributions. Eq. (\ref{eq:classifier-free-guidance}) reduces to the unconditional score when $\omega=0$, or recovers the conditional score  when $\omega=1$. 

\section{Training and Sampling Algorithms}
\label{sec:algo}

\begin{algorithm}[H]
\caption{Dual Stage Training Algorithm}
\label{algo:training}
\begin{algorithmic}[1]
    \STATE \textbf{Input:} future trajectories $\mathbf{x}_{t+1:t+\tau}$, schedule $\beta(t)$\\ 
    condition $\mathbf{c}=x_{1:t},\, c^{\text{trend}}_{t+1:t+\tau},\, c^{\text{vol}}_{t+1:t+\tau},\, c^{\text{liq}}_{t+1:t+\tau},\, c^{\text{imb}}_{t+1:t+\tau}$
    \STATE \textbf{Stage 1}
    \STATE Initialize model parameters $\theta=\{\theta_{\text{base}}, \theta_{\text{ctrl}}\}$, freeze $\theta_{\text{ctrl}}$
    \FOR{each training step}
        \STATE Sample diffusion time $t \sim \mathcal{U}(0,1)$
        \STATE Sample noise $\mathbf{z} \sim \mathcal{N}(\mathbf{0}, \mathbf{I})$
        \STATE Compute $\mathbf{f}(\mathbf{x},t), g(t)$ by $\beta(t)$
        \STATE Get perturbed data future trajectory
        \STATE Update $\theta_{\text{base}}$ by minimizing the objective in Eq. \eqref{eq:sde-training}
    \ENDFOR 
    \STATE \vspace{0.2em}
    \STATE \textbf{Stage 2}
    \STATE Freeze model parameters $\theta_{\text{base}}$, unfreeze $\theta_{\text{ctrl}}$
        \FOR{each training step}
        \STATE Sample diffusion time $t \sim \mathcal{U}(0,1)$
        \STATE Sample noise $\mathbf{z} \sim \mathcal{N}(\mathbf{0}, \mathbf{I})$
        \STATE Compute $\mathbf{f}(\mathbf{x},t), g(t)$ from the schedule
        \STATE Get perturbed data future trajectory
        \STATE Update $\theta_{\text{ctrl}}$ by minimizing the objective in Eq. \eqref{eq:sde-training}
    \ENDFOR
    \STATE \textbf{Output:} Network $\mathbf{s}_{\theta}(\mathbf{x}, t, \mathbf{c})$
\end{algorithmic}
\end{algorithm}

\begin{algorithm}[H]
\caption{Sampling Algorithm}
\label{algo:sampling}
\begin{algorithmic}[1]
    \STATE \textbf{Input:} number of steps $N$, schedule $\beta(t)$, \\ condition $\mathbf{c}=x_{1:t},\, c^{\text{trend}}_{t+1:t+\tau},\, c^{\text{vol}}_{t+1:t+\tau},\, c^{\text{liq}}_{t+1:t+\tau},\, c^{\text{imb}}_{t+1:t+\tau}$, \\
    guidance scale $w$, network $\mathbf{s}_{\theta}(\mathbf{x}, t, \mathbf{c})$
    \STATE $\mathbf{x}_1 \sim \mathcal{N}(\mathbf{0}, \mathbf{I})$
    \FOR{$i = N, N-1, \dots, 1$}
        \STATE Set diffusion time $t \leftarrow i/N$, step size $\Delta t \leftarrow 1/N$
        \STATE Sample noise $\mathbf{z} \sim \mathcal{N}(\mathbf{0}, \mathbf{I})$
        \STATE Evaluate conditional score $\mathbf{s}_c \leftarrow \mathbf{s}_{\theta}(\mathbf{x}_t, t, \mathbf{c})$
        \STATE Evaluate unconditional score $\mathbf{s}_{\emptyset} \leftarrow \mathbf{s}_{\theta}(\mathbf{x}_t, t, \varnothing)$
        \STATE Apply classifier-free guidance:
        $
        \mathbf{s} \leftarrow (1+w)\mathbf{s}_c - w\,\mathbf{s}_{\emptyset}
        $
        \STATE Update by ancestral sampling:
        $$
        \mathbf{x}_{t-\Delta t} \leftarrow \mathbf{x}_t 
        + \Big(\tfrac{1}{2}\beta(t)\mathbf{x}_t + \beta(t)\mathbf{s}\Big)\Delta t
        + \sqrt{\beta(t)\Delta t}\,\mathbf{z}
        $$
    \ENDFOR
    \STATE \textbf{Output:} generated trajectory $\hat{\mathbf{x}}_0 \leftarrow \mathbf{x}_0$
\end{algorithmic}
\end{algorithm}

\begin{figure*}[htb]  
    \centering  
    \includegraphics[width=\textwidth]{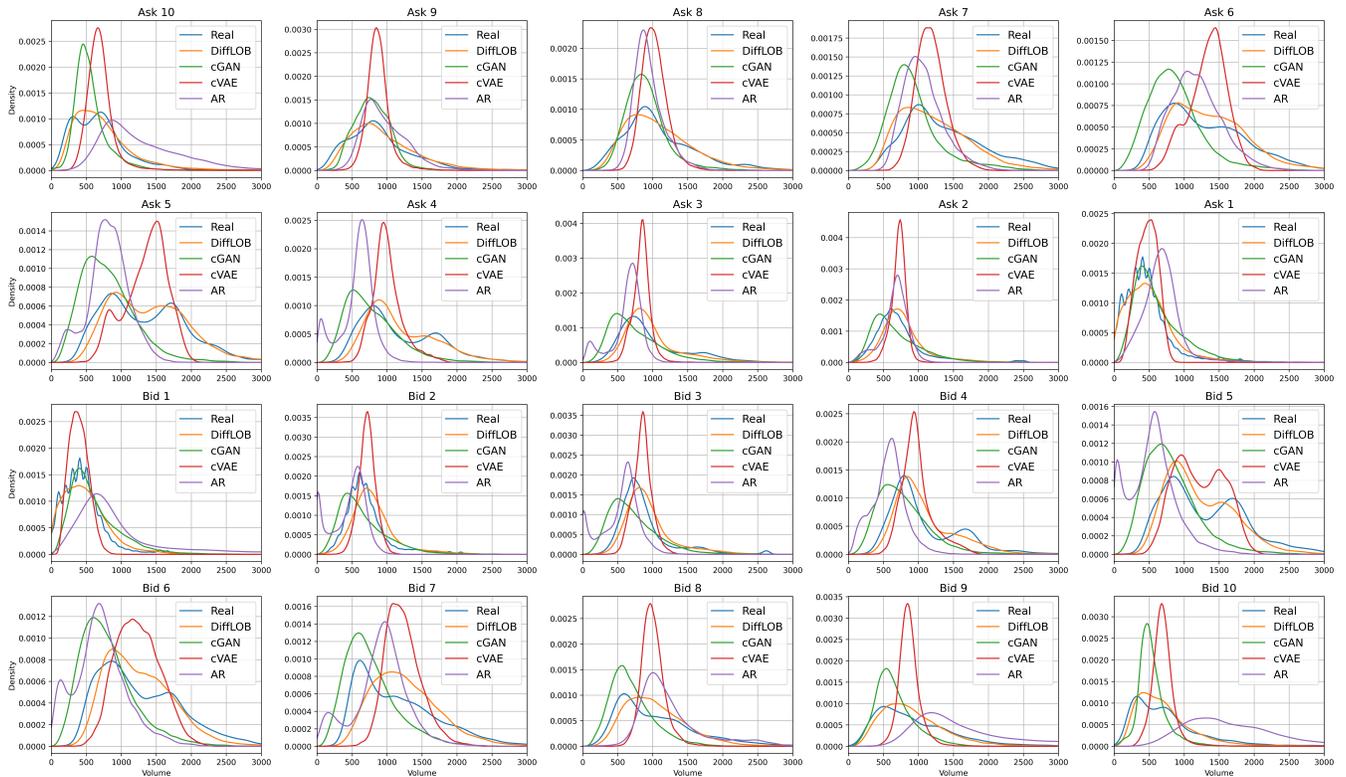}  
    \caption{Marginal Volume Distribution across Price Levels.}
    \label{fig:volume-distribution}  
\end{figure*}

We adopt a two-stage training strategy to enable stable and effective regime control. During training, conditions are randomly dropped with probability $0.5$ to enable classifier-free guidance.
In the first stage, we train the diffusion backbone together with the regime encoders while excluding the control module, allowing the model to learn the unconditional and condition-aware LOB dynamics.
In the second stage, we freeze all previously trained parameters and optimize only the control module, which injects regime-dependent intervention signals into the backbone.
This design ensures that counterfactual control is learned as a residual modification on top of a well-trained generative model, avoiding degradation of base dynamics.

At inference time, we generate LOB trajectories using an ancestral sampling procedure for the variance-preserving SDE.
Classifier-free guidance is applied during sampling by combining conditional and unconditional score estimates, enabling continuous control over the strength of regime interventions. The complete training and sampling algorithm are shown in Algorithms \ref{algo:training} and \ref{algo:sampling}, respectively.

\section{Volume Distribution}
\label{sec:volume-distribution}

Figure \ref{fig:volume-distribution} shows the marginal volume distributions across individual price levels on both the ask and bid sides. DiffLOB generally reproduces the empirical distributional shapes across depth levels, capturing both the scale and relative variation of volumes from the best quotes to deeper levels. While minor deviations remain at certain price levels, DiffLOB avoids the severe mode collapse or tail distortion observed in some baselines. In contrast, cGAN and cVAE exhibit noticeable mismatches, including shifted modes and overly concentrated densities, while the autoregressive model tends to generate excessively dispersed distributions. Overall, DiffLOB provides a more balanced approximation of marginal volume distributions across price levels.

\section{Complete Counterfactual Table}
\label{sec:complete-counterfactual-table}

Table \ref{tab:complete-counterfactual-validity} reports distributional distances between counterfactual trajectories generated under extreme regime interventions and real trajectories observed in the corresponding regimes, evaluated across three stocks. Overall, DiffLOB achieves consistently low distances across most regimes and assets, indicating stable counterfactual alignment with real market behavior. In particular, DiffLOB performs strongly under volatility, liquidity, and imbalance interventions, where it frequently attains the lowest or near-lowest Wasserstein, KL, and JS distances.

We observe that certain baselines exhibit localized strengths: cGAN achieves relatively low distances under trend interventions, while removing the control module (DiffLOB w/o C) yields competitive results in some high-liquidity and high-imbalance cases. However, these improvements are not consistent across regimes or assets. In contrast, DiffLOB maintains robust performance across different regime types and stocks, suggesting that explicit regime-aware control leads to more reliable counterfactual generation overall.

\clearpage
\begin{table*}[p]  
\scalebox{0.7}{
\begin{tabular}{@{}lrrrrrrrrrrrr@{}}
\toprule
               & \multicolumn{4}{c}{AMZN}                                                                                                                                     & \multicolumn{4}{c}{AAPL}                                                                                                                                      & \multicolumn{4}{c}{GOOG}                                                                                                                                      \\ \midrule
High-Trend     & \multicolumn{1}{c}{KS}                & \multicolumn{1}{c}{Wasserstein}       & \multicolumn{1}{c}{KL}               & \multicolumn{1}{c}{JS}                & \multicolumn{1}{c}{KS}                & \multicolumn{1}{c}{Wasserstein}       & \multicolumn{1}{c}{KL}                & \multicolumn{1}{c}{JS}                & \multicolumn{1}{c}{KS}                & \multicolumn{1}{c}{Wasserstein}       & \multicolumn{1}{c}{KL}                & \multicolumn{1}{c}{JS}                \\ \midrule
DiffLOB        & 0.444095                              & 0.522512                              & 2.517532                             & 0.180279                              & 0.527486                              & 0.705667                              & 3.79866                               & 0.228299                              & 0.319797                              & 0.38452                               & 1.943212                              & 0.130711                              \\
DiffLOB wo C   & 0.515225                              & 0.801835                              & 4.879508                             & 0.265181                              & 0.52187                               & 0.700273                              & 3.706363                              & 0.225847                              & 0.400386                              & 0.548083                              & 3.015241                              & 0.170459                              \\
Diff-CSDI      & 0.627345                              & 65.274135                             & 5.855148                             & 0.307245                              & 0.650097                              & 77.661673                             & 6.409168                              & 0.346118                              & 0.608903                              & 51.355576                             & 5.459767                              & 0.2802                                \\
Diff-S4        & 0.292354                              & 0.264849                              & 1.098681                             & 0.092509                              & 0.479872                              & 0.593335                              & 2.674102                              & 0.19118                               & 0.204638                              & 0.264558                              & 1.397245                              & 0.087298                              \\
CGAN           & \textbf{0.048718}                     & \textbf{0.036091}                     & \textbf{0.422542}                    & \textbf{0.026361}                     & 0.826926                              & 2.46174                               & 9.609552                              & 0.472167                              & 0.51297                               & 0.959163                              & 5.462742                              & 0.262343                              \\
CVAE           & 0.298339                              & 0.225797                              & 0.79771                              & 0.092942                              & 0.376703                              & 0.440777                              & 1.364931                              & 0.123675                              & 0.06105                               & 0.035715                              & 0.236427                              & 0.049655                              \\
AR             & 0.200438                              & 0.199092                              & 1.210306                             & 0.091039                              & \textbf{0.339528}                     & \textbf{0.367727}                     & \textbf{1.093274}                     & \textbf{0.098998}                     & \textbf{0.059685}                     & \textbf{0.034748}                     & \textbf{0.21401}                      & \textbf{0.034285}                     \\ \midrule
Low-Trend      & \multicolumn{1}{c}{KS}                & \multicolumn{1}{c}{Wasserstein}       & \multicolumn{1}{c}{KL}               & \multicolumn{1}{c}{JS}                & \multicolumn{1}{c}{KS}                & \multicolumn{1}{c}{Wasserstein}       & \multicolumn{1}{c}{KL}                & \multicolumn{1}{c}{JS}                & \multicolumn{1}{c}{KS}                & \multicolumn{1}{c}{Wasserstein}       & \multicolumn{1}{c}{KL}                & \multicolumn{1}{c}{JS}                \\ \midrule
DiffLOB        & \multicolumn{1}{c}{0.425788}          & \multicolumn{1}{c}{0.507738}          & \multicolumn{1}{c}{3.529521}         & \multicolumn{1}{c}{0.182637}          & \multicolumn{1}{c}{0.466021}          & \multicolumn{1}{c}{0.663595}          & \multicolumn{1}{c}{2.969223}          & \multicolumn{1}{c}{0.18528}           & \multicolumn{1}{c}{0.418144}          & \multicolumn{1}{c}{0.357243}          & \multicolumn{1}{c}{3.623893}          & \multicolumn{1}{c}{0.174686}          \\
DiffLOB wo C   & \multicolumn{1}{c}{0.339051}          & \multicolumn{1}{c}{0.36766}           & \multicolumn{1}{c}{2.525023}         & \multicolumn{1}{c}{0.141955}          & \multicolumn{1}{c}{0.372399}          & \multicolumn{1}{c}{0.52205}           & \multicolumn{1}{c}{2.179232}          & \multicolumn{1}{c}{0.141017}          & \multicolumn{1}{c}{0.389908}          & \multicolumn{1}{c}{0.326069}          & \multicolumn{1}{c}{3.413639}          & \multicolumn{1}{c}{0.161795}          \\
Diff-CSDI      & \multicolumn{1}{c}{0.755996}          & \multicolumn{1}{c}{84.376958}         & \multicolumn{1}{c}{6.563037}         & \multicolumn{1}{c}{0.356456}          & \multicolumn{1}{c}{0.703418}          & \multicolumn{1}{c}{93.138326}         & \multicolumn{1}{c}{6.243724}          & \multicolumn{1}{c}{0.337338}          & \multicolumn{1}{c}{0.598181}          & \multicolumn{1}{c}{45.326588}         & \multicolumn{1}{c}{5.281554}          & \multicolumn{1}{c}{0.267412}          \\
Diff-S4        & \multicolumn{1}{c}{0.431509}          & \multicolumn{1}{c}{0.521119}          & \multicolumn{1}{c}{3.855445}         & \multicolumn{1}{c}{0.187511}          & \multicolumn{1}{c}{0.226887}          & \multicolumn{1}{c}{0.342775}          & \multicolumn{1}{c}{1.410633}          & \multicolumn{1}{c}{0.094289}          & \multicolumn{1}{c}{0.43068}           & \multicolumn{1}{c}{0.359478}          & \multicolumn{1}{c}{3.91257}           & \multicolumn{1}{c}{0.192965}          \\
CGAN           & \multicolumn{1}{c}{\textbf{0.039677}} & \multicolumn{1}{c}{\textbf{0.044298}} & \multicolumn{1}{c}{\textbf{0.47366}} & \multicolumn{1}{c}{\textbf{0.023752}} & \multicolumn{1}{c}{0.821983}          & \multicolumn{1}{c}{2.436247}          & \multicolumn{1}{c}{9.382776}          & \multicolumn{1}{c}{0.462602}          & \multicolumn{1}{c}{0.267701}          & \multicolumn{1}{c}{0.179529}          & \multicolumn{1}{c}{1.760093}          & \multicolumn{1}{c}{0.108268}          \\
CVAE           & \multicolumn{1}{c}{0.38066}           & \multicolumn{1}{c}{0.459526}          & \multicolumn{1}{c}{2.89087}          & \multicolumn{1}{c}{0.156796}          & \multicolumn{1}{c}{\textbf{0.192543}} & \multicolumn{1}{c}{\textbf{0.262382}} & \multicolumn{1}{c}{\textbf{1.078224}} & \multicolumn{1}{c}{\textbf{0.074031}} & \multicolumn{1}{c}{0.299508}          & \multicolumn{1}{c}{0.201727}          & \multicolumn{1}{c}{1.934686}          & \multicolumn{1}{c}{0.122522}          \\
AR             & \multicolumn{1}{c}{0.264977}          & \multicolumn{1}{c}{0.296113}          & \multicolumn{1}{c}{1.946188}         & \multicolumn{1}{c}{0.107689}          & \multicolumn{1}{c}{0.365958}          & \multicolumn{1}{c}{0.535547}          & \multicolumn{1}{c}{2.296313}          & \multicolumn{1}{c}{0.139862}          & \multicolumn{1}{c}{\textbf{0.182677}} & \multicolumn{1}{c}{\textbf{0.110635}} & \multicolumn{1}{c}{\textbf{0.701304}} & \multicolumn{1}{c}{\textbf{0.061796}} \\ \midrule
High-Volatilty & \multicolumn{1}{c}{KS}                & \multicolumn{1}{c}{Wasserstein}       & \multicolumn{1}{c}{KL}               & \multicolumn{1}{c}{JS}                & \multicolumn{1}{c}{KS}                & \multicolumn{1}{c}{Wasserstein}       & \multicolumn{1}{c}{KL}                & \multicolumn{1}{c}{JS}                & \multicolumn{1}{c}{KS}                & \multicolumn{1}{c}{Wasserstein}       & \multicolumn{1}{c}{KL}                & \multicolumn{1}{c}{JS}                \\ \midrule
DiffLOB        & \textbf{0.083474}                     & \textbf{0.06075}                      & \textbf{0.169118}                    & \textbf{0.03113}                      & 0.195453                              & 0.170572                              & 0.390789                              & 0.063731                              & 0.208521                              & 0.162132                              & 0.958381                              & 0.114761                              \\
DiffLOB wo C   & 0.292613                              & 0.237796                              & 1.004272                             & 0.108261                              & 0.212819                              & 0.188072                              & \textbf{0.31742}                      & 0.062189                              & 0.276429                              & 0.18771                               & \textbf{0.753739}                     & \textbf{0.099716}                     \\
Diff-CSDI      & 0.609586                              & 59.726185                             & 5.559404                             & 0.286991                              & 0.585045                              & 67.233425                             & 5.500033                              & 0.286357                              & 0.573925                              & 46.365704                             & 5.005716                              & 0.252426                              \\
Diff-S4        & 0.118421                              & 0.12436                               & 0.67194                              & 0.064088                              & 0.309172                              & 0.287524                              & 0.65901                               & 0.086488                              & 0.242306                              & 0.202847                              & 1.323483                              & 0.142103                              \\
CGAN           & 0.178244                              & 0.088973                              & 0.524627                             & 0.048186                              & 0.813774                              & 2.573336                              & 9.687475                              & 0.478825                              & 0.483661                              & 0.512713                              & 2.530705                              & 0.195179                              \\
CVAE           & 0.118761                              & 0.077975                              & 0.646276                             & 0.071165                              & 0.277909                              & 0.230906                              & 0.763321                              & 0.086946                              & \textbf{0.204925}                     & 0.180972                              & 1.319116                              & 0.140161                              \\
AR             & 0.083668                              & 0.065337                              & 0.454682                             & 0.041336                              & \textbf{0.108995}                     & \textbf{0.103518}                     & 0.617903                              & \textbf{0.061673}                     & 0.208233                              & \textbf{0.161}                        & 0.85236                               & 0.120365                              \\ \midrule
Low-Volatility & \multicolumn{1}{c}{KS}                & \multicolumn{1}{c}{Wasserstein}       & \multicolumn{1}{c}{KL}               & \multicolumn{1}{c}{JS}                & \multicolumn{1}{c}{KS}                & \multicolumn{1}{c}{Wasserstein}       & \multicolumn{1}{c}{KL}                & \multicolumn{1}{c}{JS}                & \multicolumn{1}{c}{KS}                & \multicolumn{1}{c}{Wasserstein}       & \multicolumn{1}{c}{KL}                & \multicolumn{1}{c}{JS}                \\ \midrule
DiffLOB        & \textbf{0.150047}                     & \textbf{0.072569}                     & \textbf{0.772114}                    & \textbf{0.061085}                     & 0.263016                              & 0.16332                               & 1.228604                              & \textbf{0.080622}                     & 0.171852                              & 0.071879                              & 1.067833                              & 0.095288                              \\
DiffLOB wo C   & 0.163749                              & 0.140627                              & 1.38697                              & 0.075512                              & 0.197904                              & 0.12845                               & 1.387818                              & 0.087776                              & 0.201614                              & 0.089192                              & 1.05244                               & \textbf{0.081389}                     \\
Diff-CSDI      & 0.636904                              & 59.84074                              & 5.663091                             & 0.293081                              & 0.634541                              & 66.696715                             & 5.392071                              & 0.279371                              & 0.56042                               & 45.873914                             & 5.429973                              & 0.277177                              \\
Diff-S4        & 0.297515                              & 0.160111                              & 0.999069                             & 0.111148                              & \textbf{0.19326}                      & \textbf{0.126173}                     & \textbf{1.192786}                     & 0.086189                              & 0.190067                              & 0.080766                              & 1.689373                              & 0.118913                              \\
CGAN           & 0.16065                               & 0.113201                              & 1.285274                             & 0.078348                              & 0.850427                              & 2.242279                              & 9.860524                              & 0.47597                               & 0.353784                              & 0.206637                              & 1.998859                              & 0.146298                              \\
CVAE           & 0.287089                              & 0.16455                               & 1.092421                             & 0.100044                              & 0.297702                              & 0.188888                              & 1.299558                              & 0.097106                              & 0.149426                              & 0.08095                               & 1.107241                              & 0.098775                              \\
AR             & 0.244042                              & 0.141                                 & 1.457307                             & 0.09852                               & 0.362483                              & 0.292242                              & 1.783818                              & 0.118306                              & \textbf{0.124769}                     & \textbf{0.070454}                     & \textbf{0.925041}                     & 0.092233                              \\ \midrule
High-Liquidity & \multicolumn{1}{c}{KS}                & \multicolumn{1}{c}{Wasserstein}       & \multicolumn{1}{c}{KL}               & \multicolumn{1}{c}{JS}                & \multicolumn{1}{c}{KS}                & \multicolumn{1}{c}{Wasserstein}       & \multicolumn{1}{c}{KL}                & \multicolumn{1}{c}{JS}                & \multicolumn{1}{c}{KS}                & \multicolumn{1}{c}{Wasserstein}       & \multicolumn{1}{c}{KL}                & \multicolumn{1}{c}{JS}                \\ \midrule
DiffLOB        & 0.137538                              & \textbf{168.024319}                   & 0.208135                             & 0.034988                              & \textbf{0.123554}                     & \textbf{93.884826}                    & \textbf{0.149982}                     & \textbf{0.025053}                     & \textbf{0.150165}                     & \textbf{184.78402}                    & 0.204472                              & \textbf{0.033355}                     \\
DiffLOB wo C   & \textbf{0.134555}                     & 179.447363                            & \textbf{0.18559}                     & \textbf{0.03162}                      & 0.127951                              & 105.880109                            & 0.171677                              & 0.02943                               & 0.161624                              & 194.000928                            & \textbf{0.203039}                     & 0.036367                              \\
Diff-CSDI      & 0.318472                              & 617445.7056                           & 0.419245                             & 0.05283                               & 0.265748                              & 611953.9495                           & 0.303013                              & 0.034818                              & 0.282669                              & 592234.437                            & 0.412651                              & 0.040742                              \\
Diff-S4        & 0.199552                              & 229.470802                            & 0.279902                             & 0.050367                              & 0.154274                              & 137.216909                            & 0.22666                               & 0.034013                              & 0.239978                              & 290.128272                            & 0.467916                              & 0.068077                              \\
CGAN           & 0.389481                              & 513.984844                            & 1.1278                               & 0.153755                              & 0.629324                              & 407.898305                            & 4.29376                               & 0.320787                              & 0.222264                              & 312.509538                            & 0.411387                              & 0.064169                              \\
CVAE           & 0.335492                              & 365.126803                            & 1.982426                             & 0.229481                              & 0.327466                              & 238.102708                            & 2.251998                              & 0.219447                              & 0.262501                              & 315.088815                            & 1.339358                              & 0.153376                              \\
AR             & 0.422048                              & 445.418425                            & 1.840147                             & 0.215084                              & 0.339363                              & 246.09984                             & 1.292155                              & 0.159703                              & 0.363537                              & 398.86276                             & 1.221241                              & 0.169614                              \\ \midrule
Low-Liquidity  & \multicolumn{1}{c}{KS}                & \multicolumn{1}{c}{Wasserstein}       & \multicolumn{1}{c}{KL}               & \multicolumn{1}{c}{JS}                & \multicolumn{1}{c}{KS}                & \multicolumn{1}{c}{Wasserstein}       & \multicolumn{1}{c}{KL}                & \multicolumn{1}{c}{JS}                & \multicolumn{1}{c}{KS}                & \multicolumn{1}{c}{Wasserstein}       & \multicolumn{1}{c}{KL}                & \multicolumn{1}{c}{JS}                \\ \midrule
DiffLOB        & \textbf{0.126894}                     & \textbf{104.19346}                    & \textbf{0.099957}                    & \textbf{0.021029}                     & \textbf{0.084143}                     & \textbf{42.026082}                    & \textbf{0.069565}                     & \textbf{0.015318}                     & \textbf{0.092995}                     & \textbf{93.784744}                    & \textbf{0.085606}                     & \textbf{0.01421}                      \\
DiffLOB wo C   & 0.140489                              & 110.790344                            & 0.114877                             & 0.025278                              & 0.107366                              & 59.18505                              & 0.09517                               & 0.021885                              & 0.119772                              & 114.837243                            & 0.092891                              & 0.018651                              \\
Diff-CSDI      & 0.311335                              & 585813.0602                           & 0.444758                             & 0.073959                              & 0.241414                              & 583028.0427                           & 0.263493                              & 0.046031                              & 0.2421                                & 563904.5316                           & 0.257089                              & 0.045014                              \\
Diff-S4        & 0.224048                              & 142.265977                            & 0.212572                             & 0.048326                              & 0.156323                              & 100.205219                            & 0.146806                              & 0.030725                              & 0.284745                              & 264.630407                            & 0.379443                              & 0.077248                              \\
CGAN           & 0.214546                              & 167.358666                            & 0.315572                             & 0.063397                              & 0.405366                              & 194.850894                            & 2.043172                              & 0.201632                              & 0.220858                              & 195.332252                            & 0.213879                              & 0.047721                              \\
CVAE           & 0.351886                              & 231.117557
                            & 1.748266
                             & 0.220437
                               & 0.292353
                              & 148.81207
                            & 1.759316
                              & 0.184381
                              & 0.286505
                              & 214.83184
                           & 1.315952
        & 0.155636
                              \\
AR             & 0.410027                              & 274.375898                            & 1.572332                             & 0.19979                               & 0.318566                              & 162.614041                            & 1.41773                               & 0.178782                              & 0.314595                              & 235.733495                            & 1.105514                              & 0.145604                              \\ \midrule
High-Imbalance & \multicolumn{1}{c}{KS}                & \multicolumn{1}{c}{Wasserstein}       & \multicolumn{1}{c}{KL}               & \multicolumn{1}{c}{JS}                & \multicolumn{1}{c}{KS}                & \multicolumn{1}{c}{Wasserstein}       & \multicolumn{1}{c}{KL}                & \multicolumn{1}{c}{JS}                & \multicolumn{1}{c}{KS}                & \multicolumn{1}{c}{Wasserstein}       & \multicolumn{1}{c}{KL}                & \multicolumn{1}{c}{JS}                \\ \midrule
DiffLOB        & 0.142787                              & \textbf{154.156402}                   & 0.182522                             & 0.032153                              & \textbf{0.096136}                     & \textbf{78.177124}                    & \textbf{0.111676}                     & \textbf{0.020894}                     & \textbf{0.114366}                     & \textbf{109.782946}                   & \textbf{0.122575}                     & \textbf{0.025946}                     \\
DiffLOB wo C   & \textbf{0.130096}                     & 155.84261                             & 0.181721                    & \textbf{0.032042}                     & 0.122496                              & 102.745785                            & 0.166532                              & 0.028015                              & 0.13587                               & 140.918735                            & 0.163554                              & 0.03445                               \\
Diff-CSDI      & 0.303622                              & 629928.7738                           & 0.363013                             & 0.054523                              & 0.261696                              & 584543.7943                           & 0.267411                              & 0.037983                              & 0.28097                               & 605229.5742                           & 0.420215                              & 0.0584                                \\
Diff-S4        & 0.163933                              & 165.329592                            & \textbf{0.180472}                             & 0.035623                              & 0.151473                              & 119.539495                            & 0.181243                              & 0.032926                              & 0.233546                              & 307.169842                            & 0.244487                              & 0.046097                              \\
CGAN           & 0.285831                              & 342.915205                            & 0.59001                              & 0.095159                              & 0.484773                              & 321.149662                            & 2.998772                              & 0.247907                              & 0.206432                              & 232.264564                            & 0.237424                              & 0.045062                              \\
CVAE           & 0.274923                              & 291.958093                            & 1.485293                             & 0.171144                              & 0.27594                               & 206.120915                            & 1.70025                               & 0.170524                              & 0.272183                              & 284.282626                            & 1.338171                              & 0.156963                              \\
AR             & 0.432061                              & 397.430331                            & 1.656605                             & 0.202636                              & 0.309476                              & 211.81121                             & 1.028418                              & 0.141033                              & 0.342737                              & 339.033016                            & 1.231251                              & 0.16067                               \\ \midrule
Low-Imbalance  & \multicolumn{1}{c}{KS}                & \multicolumn{1}{c}{Wasserstein}       & \multicolumn{1}{c}{KL}               & \multicolumn{1}{c}{JS}                & \multicolumn{1}{c}{KS}                & \multicolumn{1}{c}{Wasserstein}       & \multicolumn{1}{c}{KL}                & \multicolumn{1}{c}{JS}                & \multicolumn{1}{c}{KS}                & \multicolumn{1}{c}{Wasserstein}       & \multicolumn{1}{c}{KL}                & \multicolumn{1}{c}{JS}                \\ \midrule
DiffLOB        & \textbf{0.138717}                     & \textbf{147.853412}                   & \textbf{0.181154}                    & \textbf{0.034248}                     & \textbf{0.139123}                     & \textbf{81.910726}                    & \textbf{0.133491}                     & \textbf{0.025795}                     & \textbf{0.103865}                     & \textbf{100.254673}                   & \textbf{0.170693}                     & \textbf{0.026857}                     \\
DiffLOB wo C   & 0.153997                              & 174.322966                            & 0.204804                             & 0.038629                              & 0.144674                              & 88.387008                             & 0.147903                              & 0.031433                              & 0.130511                              & 131.0986                              & 0.178989                              & 0.032156                              \\
Diff-CSDI      & 0.307549                              & 593616.0608                           & 0.43386                              & 0.063705                              & 0.217315                              & 577551.335                            & 0.259399                              & 0.04002                               & 0.298334                              & 673911.8064                           & 0.46533                               & 0.074212                              \\
Diff-S4        & 0.198524                              & 209.414158                            & 0.253176                             & 0.049452                              & 0.162106                              & 130.026248                            & 0.229766                              & 0.033792                              & 0.268451                              & 285.821643                            & 0.493237                              & 0.07783                               \\
CGAN           & 0.298427                              & 356.49669                             & 0.635989                             & 0.101998                              & 0.461362                              & 250.252468                            & 2.855038                              & 0.22801                               & 0.226664                              & 242.115669                            & 0.276046                              & 0.05371                               \\
CVAE           & 0.290173                              & 290.269725                            & 1.529841                             & 0.18089                               & 0.278539                              & 180.582864                            & 1.551757                              & 0.173447                              & 0.266441                              & 253.438419                            & 1.22087                               & 0.142463                              \\
AR             & 0.383488                              & 338.46483                             & 1.580578                             & 0.186188                              & 0.310076                              & 190.853724                            & 1.075461                              & 0.154077                              & 0.32443                               & 297.557395                            & 1.323793                              & 0.171135                              \\ \bottomrule
\end{tabular}}
\caption{Counterfactual Validity on the three Stocks.}
\label{tab:complete-counterfactual-validity}
\end{table*} 

\end{document}